\documentclass[10pt,a4paper,final]{iopart}

\usepackage{amssymb}
\usepackage{graphicx}
\usepackage{ulem}
\usepackage[breaklinks=true,colorlinks=true,linkcolor=blue,urlcolor=blue,citecolor=blue]{hyperref}

\begin{document}
\title{Regularity and chaos in cavity QED}

\author{Miguel Angel Bastarrachea-Magnani$^{1,2}$\footnote{On postdoctoral leave from Instituto de Ciencias Nucleares, Universidad Nacional Aut\'onoma de M\'exico}, Baldemar L\'opez-del-Carpio$^3$, Jorge Ch\'avez-Carlos$^2$, Sergio Lerma-Hern\'andez$^{2,3}$\footnote{On sabbatical leave from Facultad de F\'isica, Universidad Veracruzana.} and Jorge G. Hirsch$^2$\footnote{Correspondence author.}.}
\address{$^1$ Physikalisches Institut, Albert-Ludwigs-Universitat Freiburg, Hermann-Herder-Str. 3, Freiburg, Germany, D-79104.}
\address{$^2$ Instituto de Ciencias Nucleares, Universidad Nacional Aut\'onoma de M\'exico,
Apdo. Postal 70-543, Cd. Mx., M\'exico, C.P. 04510}
\address{$^3$ Facultad de F\'\i sica, Universidad Veracruzana,
Circuito Aguirre Beltr\'an s/n, Xalapa, Veracruz, M\'exico, C.P. 91000}

\ead{miguel.bastarrachea@nucleares.unam.mx, ramedlab1@gmail.com, fis$\_$jorge.chavez@yahoo.com.mx, slerma@uv.mx, and hirsch@nucleares.unam.mx}

\begin{abstract}
The interaction of a quantized electromagnetic field in a cavity with a set of two-level atoms inside it can be described with algebraic Hamiltonians of
increasing complexity, from the Rabi to the Dicke models. Their algebraic character allows, through the use of coherent states, a semiclassical description in phase space, where the non-integrable Dicke model has regions associated with regular and chaotic motion.  The appearance of classical chaos can be quantified calculating the largest Lyapunov exponent along the whole available phase space for a given energy. In the quantum regime, employing efficient diagonalization techniques, we are able to perform a detailed quantitative study of the regular and chaotic regions, where the quantum participation ratio ($P_R$) of coherent states on the eigenenergy basis plays a role equivalent to the Lyapunov exponent. 
It is noted that, in the thermodynamic limit, dividing the participation ratio by the number of atoms leads to a positive value in chaotic regions, while it tends to zero in the regular ones.
\end{abstract}

\vspace{2pc}

\submitto{\PS}

\section{Introduction}

The problem of how isolated quantum many-body systems could attain a state of equilibrium and its dynamical implications in the field of quantum thermodynamics, has become a fundamental issue where many challenging questions remain open \cite{Pol11,Eis15,Gog15,Aless16}. It has been shown that in a quantum system equilibration can occur for unitary dynamics under conditions of classical chaos \cite{Alt12}. Therefore, quantum chaos or the study of quantum system which have a classical chaotic counterpart \cite{Berry,Bohigas,BerryRobnik}, has become a very important topic in this context  \cite{Par89,Gutz,Stro94,Haake01}. Typically, classical many-body systems containing a big number of degrees of freedom tend to have chaos as they are non-integrable, i.e. they do not have as much conserved quantities as degrees of freedom. In the classical realm, non-integrability, chaos and ergodicity are intimately related. However, these definitions fail to capture the quantum realm due to the lack of trajectories in the quantum phase space.

Algebraic Hamiltonians provide a natural scenario to explore these subjects. The algebra gives a simple way to build coherent states, which are considered the most classical quantum states. Coherent states provide a simple connection between quantum and classical realm.
The Dicke model \cite{Dicke54} is described by an algebraic Hamiltonian which has gained renewed interest in the last years thanks to its combination of algebraic simplicity and its rich phenomenology. It was meant to describe a collective system of $\mathcal{N}$ atoms within the two-level approximation interacting with a single mode of radiation field inside a QED cavity. However, its algebraic simplicity makes it suitable for the description of systems of interacting qubits within the quantum information framework. From Bose-Einstein condensates in optical lattices, to polaritons, and superconducting qubits in circuit QED, the Dicke model has become a paradigmatic example of quantum collective behavior \cite{Sche03,Blais04,Sche07,Fink09,Bau10,Nag10,Nie10,Cas10,Mezza14}. 

As a toy model, the Dicke Hamltonian is a good start for several approximated Hamiltonians and for different approaches to describe quantum systems. Particularly, the model is well-known thanks to its superradiant thermal phase transition \cite{HL73,WH73,CGW73,CD74,Basta16JS} that persists in the zero-temperature, giving rise to a quantum phase transition (QPT) \cite{Emary03}. In connection to the ground-state QPT it has other interesting features like entanglement \cite{Lam05,Vid06}, and
excited-state quantum phase transitions (ESQPT) \cite{Cej06,Cap08,Str09,Str14,Str16}. It is particularly relevant for this work that it presents quantum chaos. 
The Dicke model describes one of the simplest nonintegrable atom-field system, exhibiting quantum chaos. In the Dicke model the evolution equation of the Husimi function is of the Fokker-Planck type, and the transition from ergodic to non-ergodic behavior can be quantified employing the average over an energy shell of the inverse participation ratio between the eigenstates of the integrable and the total Hamiltonian. In this way the thermal behavior of the system is closely related with delocalization\cite{Cano11,Pal10}. Thus, thanks to its simplicity the Dicke model opens a door for the study of equilibrium and thermalization in isolated quantum many-body systems, as well as for the classical-quantum correspondence.

In this comment we present a description of recent methods developed by our group to gauge the presence of chaos in quantum systems described by algebraic Hamiltonians. The qualitative description of regularity and chaos in the Dicke model has been performed employing  Poincar\'e sections for the classical analysis and Peres Lattices in the quantum case. They have been analyzed in detail in our previous comment \cite{Basta15PS}. Also, in this contribution we review the study of the classical dynamics employing Lyapunov exponents as presented in \cite{Cha16}, as well as its comparison with the quantum participation ratio discussed in \cite{Basta16PRE}, providing a self-contained introduction to the subject, and including new results which strengthen the conclusion that, in algebraic systems, the quantum participation ratio ($P_R$) of coherent states on the eigenenergy basis plays a role equivalent to the Lyapunov exponent. 

The article is organized as follows. In section 2 we describe the Dicke Hamiltonian and its classical limit. In section 3 we review the participation ratio concept and illustrate the use of the $P_{R}$ of coherent states on the Hamiltonian eigenbasis in a simple integrable algebraic model, the Lipkin-Meshkov-Glick model. Afterwards, in section 4 it is used to quantify chaos in phase space using the eigenstates of the Dicke Hamiltonian. In section 5 we discuss the scaling of the participation ratio. Finally, we expose our conclusions. 

\section{Quantum and classical Dicke Hamiltonians}

The Dicke Hamiltonian is written in terms of the operators of the Heisenberg-Weyl $HW_{4}$ and the $SU(2)$ algebras, for the boson and collective pseudospin parts, respectively. It reads (with $\hbar=1$) 
\begin{equation}
\hat{H}_{D}=\omega \hat{a}^{\dagger}\hat{a}+\omega_{0}\hat{J}_{z}+\frac{2\gamma}{\sqrt{\mathcal{N}}}\left(\hat{a}+\hat{a}^{\dagger}\right)\hat{J}_{x}.
\end{equation}
The $\hat{a}^{\dagger}$ and $\hat{a}$ are the creation and annihilation operators of the boson field, $\hat{J}_{z}$, $\hat{J}_{x}$, $\hat{J}_{y}$ are collective atomic pseudo-spin operators, and $j(j+1)$ is the eigenvalue of $\hat{\mathbf{J}}^{2}=\hat{J}_{x}^{2}+\hat{J}_{y}^{2}+\hat{J}_{z}^{2}$, where $j$ is the pseudospin length. The parameters of the Hamiltonian are $\omega$, $\omega_{0}$, and $\gamma$, which correspond to the boson frequency, the two-level energy splitting, and the coupling between the boson and the pseudospin. 
In addition, the Hamiltonian has two symmetries. First, it commutes with the $\hat{\mathbf{J}}^{2}$ operator, which divides the Hilbert space into subspaces with fixed $j$. In the last years the study of the Hamiltonian properties has been restricted to the subspace $j=\mathcal{N}/2$ because it corresponds to the maximum value of the pseudospin length and defines the symmetric atomic subspace which includes the ground state of the whole system. The second symmetry is related to the parity operator $\hat{\Pi}=e^{i\pi\hat{\Lambda}}$, with $\hat{\Lambda}=\hat{a}^{\dagger}\hat{a}+\hat{J}_{z}+j$. The eigenvalues $\lambda=n+m+j$ of the $\hat{\Lambda}$ operator are the total number of excitations, where $n$ is the number of photons and $n_{exc}=m+j$ the number of excited atoms. As the Hamiltonian commutes with the parity operator, its eigenstates have one of the two different parities ($p=\pm$). 

In spite of the presence of the parity as a second conserved quantity, the Hamiltonian has not enough conserved quantities as degrees of freedom to label every single eigenstate with quantum numbers associated to a complete set of commuting operators. In this sense, the Dicke Hamiltonian is non-integrable, and thus, it presents quantum chaos signatures. On the other hand, within the rotating-wave approximation (RWA) the Dicke model becomes the Tavis-Cummings model \cite{TC68}, which commutes with $\hat{\Lambda}$, making itself integrable. 

\subsection{Classical Hamiltonian}

By employing the coherent states of the $HW_{4}$ and $SU(2)$ algebra, i.e., the Glauber and Bloch coherent states, it is possible to obtain an effective classical Hamiltonian  \cite{MAM91}. 
Even though it is not the only way to get a correspondent classical Dicke Hamiltonian \cite{Gra84}, the one obtained via this coherent state representation has proved to be very useful  \cite{OCasta11,OCasta11a,Basta14PRA1,Basta14PRA2,Basta15PS,Basta16PRE,Cha16}. The validity of the use of coherent states is under the assumption that the system remains in the coherent state product during the temporal evolution, so its dynamical properties are described mainly by it \cite{Bak13}. Alternatively, the classical Hamiltonian obtained by considering the expectation value of the Dicke Hamiltonian in the coherent state product, is the resulting lowest order semiclassical approximation to the quantum propagator written in terms of coherent states \cite{Ribeiro06}.   

We calculate the expectation value of the Hamiltonian in the coherent state product $|\alpha\rangle\otimes |z\rangle$ of Glauber and Bloch coherent states \cite{MAM91}, defined as 
\begin{eqnarray}
|\alpha\rangle&=e^{-|\alpha|^2/2}e^{\alpha a^\dagger}|0\rangle,\\
|z\rangle&=\frac{1}{\left(1+\left|z\right|^{2}\right)^{j}} e^{z J_+}|j, -j\rangle.
\label{cs}
\end{eqnarray}
A classical description with canonical variables  is obtained from the complex parameters $\alpha$ and $z$ defining $\alpha=\sqrt{\frac{j}{2}}(q+i p)$ with $q$ and $p$  real, and using the stereographic projection of $z=\tan(\theta/2)e^{i \phi }$, from which we define 
$\tilde{j_z}=(j_z/j)=-\cos\theta$ and $\phi=\arctan(j_y/j_x)$, where $\theta$ and $\phi$ are spherical angular variables of a classical vector $\vec{j}=(j_x,j_y,j_z)$  with $|\vec{j}|=j$, and $\theta$ measured respect to the negative $z$-axis. The classical Hamiltonian per particle reads
\begin{eqnarray}
&h_{cl}(p,q,\tilde{j_z},\phi)=\frac{\langle \alpha, z| H_D|\alpha, z\rangle}{j}\\
&=\omega_{0}\tilde{j_{z}}+\frac{\omega}{2}\left(q^{2}+p^{2}\right)+2\gamma \sqrt{1- \tilde{j_{z}}^{2} }\,q\,\cos \phi,
\label{hacl}
\end{eqnarray}
with classical equations of motion given by 
$$
dq/dt=\partial h_{cl}/\partial p \ \ \ \ dp/dt=-\partial h_{cl}/\partial q
$$
$$
d\phi/dt=\partial h_{cl}/\partial \tilde{j}_z \ \ \ \  d\tilde{j}_z/dt=-\partial h_{cl}/\partial \phi.
$$
The set of equations which must be solved to calculate the Lyapunov exponents can be found in \cite{Basta14PRA1}.

\subsection{Parameters Space and QPT}

As mentioned before, the Dicke Hamiltonian presents Quantum Phase Transitions (QPT), both in the ground-state and in the spectrum, the so-called Excited-State Quantum Phase Transitions (ESQPT). 
The ground-state QPT has been widely studied along the years. It describes the transition from a normal to a superradiant phase, and has been classified as a second order phase transition in the thermodynamical sense \cite{HL73,WH73,CGW73,CD74}.
When the coupling reaches a critical value $\gamma_{c}=\sqrt{\omega\omega_{0}}/2$ a macroscopic population of the upper atomic level takes place, in which the number of photons and excited atoms scales with $\mathcal{N}$, i.e. a superradiant state appears in the ground state of the system. This separates the parameter space in two parts: the normal phase ($\gamma<\gamma_{c}$), with no photons and no excited atoms, and the superradiant phase ($\gamma>\gamma_{c}$). Given that the number of bosons is unbounded,  the range of energies $\epsilon=E/j$ is only lower bounded. The second-order QPT appears as a discontinuity in the second derivative of the ground-state energy $\epsilon_{0}(\gamma)$. It can be calculated semi-classically \cite{OCasta11,OCasta09,Nah13}, and  can be expressed as 
 \begin{equation}
\epsilon_{0}(\gamma)=\left\{\begin{array}{lr} -\omega_0  & {\hbox{for }} \gamma\leq \gamma_c,  \\
-\frac{\omega_0 }{2} \left(\frac{\gamma_c^2}{\gamma^2}+ \frac{\gamma^2}{\gamma_c^2}\right) &{\hbox{for }} \gamma > \gamma_c . \\ \end{array} \right. 
\label{eq:gse}
\end{equation}

\begin{figure}
\centering
\begin{tabular}{c}
\includegraphics[width=0.95\textwidth]{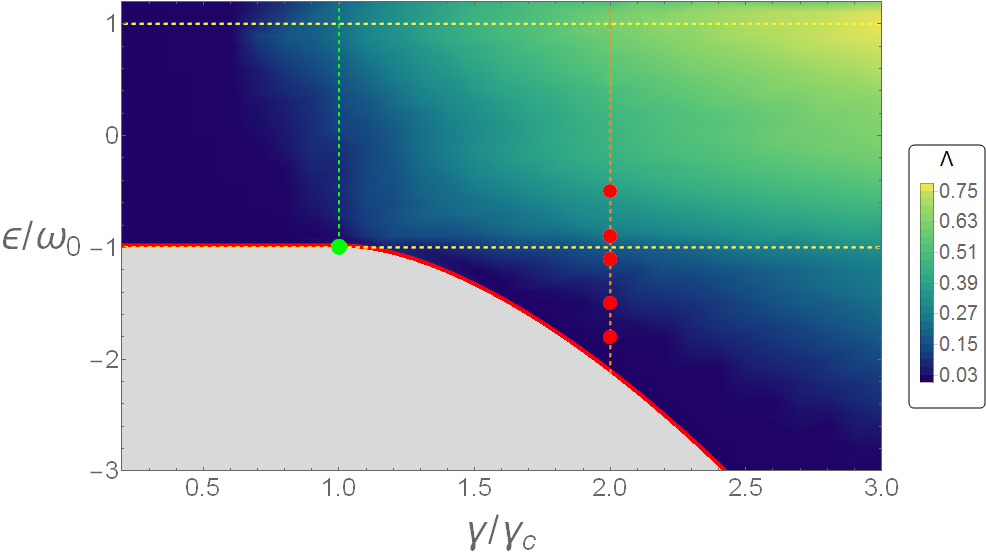} 
\end{tabular} 
\caption{ Averaged Lyapunov exponent over constant energy shells in the energy-coupling space, for  resonance $\omega=\omega_0=1$. Dark blue areas correspond to a zero Lyapunov exponent, thus are regular, meanwhile light green ones have values different from zero and are chaotic. The red points indicate the cases where the $P_{R}$ is compared against the Lyapunov exponent. These cases have the coupling $\gamma=2.0\gamma_{c}$ and energies $E/(j\omega_0)=\epsilon/\omega_0=-1.8,-1.5,-1.1,-0.9$ and $-0.5$. Thick red line indicates the ground-state energy, whereas the green dot signals the QPT and the vertical dashed green line separates the space between the normal and the superradiant phases. Finally, the dashed horizontal yellow lines signal the ESQPT energies at $\epsilon/\omega_o=-1$ and $1$.}
\label{fig:1}
\end{figure}

On the other hand, as we go up in energy in the superradiant region, the energy surface of the semiclassical Hamiltonian exhibits different structures in connection with changes in the available phase space. These changes are related to fixed points and produce singularities in the slope of the quantum Density of States (DoS), called Excited-State Quantum Phase Transitions \cite{Str09,Str14,Str16,Basta15PS,Basta16PRE,Per11,Bran13,Basta14PRA1,Basta14PRA2,Cej16,Kloc16}. While the QPT separates the parameter space of the Hamiltonian in two phases, the ESQPT separates the energy-parameter space into three regions. In the interval $\epsilon \in [\epsilon_{0}(\gamma),-\omega_0 ]$, the classical surface of constant energy is formed by two disconnected lobes (see top row of Fig.\ref{fig:4}). At  $\epsilon=-\omega_0 $ both lobes get connected indicating the first ESQPT, which we have called {\em dynamical} as it is only present in the superradiant phase, i.e. it depends on the parameters of the Hamiltonian. The second energy region is in the interval $\epsilon \in [-\omega_0 ,+\omega_0 ]$. There, the energy surface corresponds to a single lobe restricted to a limited region inside the Bloch sphere (see bottom row of Fig.\ref{fig:4}). Finally, a third region appears for energies greater than  $\omega_0 $. Here, the entire Bloch sphere is available, saturating the atomic phase space. The last two regions exist in both the normal and superradiant phases and they are separated by a second ESQPT which remains independent of the coupling, so we called it \emph{static} \cite{Basta15PS,Basta16PRE,Bran13,Basta14PRA1,Basta14PRA2}.
The Dicke Hamiltonian is also very interesting because it possess regions with mixed regularity and chaos. At first, it was thought the ESQPT was connected with the onset of chaos. However, recent works have shown that chaos and ESQPTs are two distinct phenomena whose relation depends on the system \cite{Str14,Str16,Cha16}. A map of chaos and regularity can be done by calculating the averaged Lyapunov exponent over constant energy shells in the energy-parameter space. This map gives us insight about the chaotic behavior of the Hamiltonian and how it is related with the quantum version \cite{Cha16}. In Fig.\ref{fig:1} we show this map, as well as the five points where we are going to compare the Lyapunov exponent with the quantum measure of classical chaos we propose. 

The Lyapunov exponents are asymptotic measures characterizing the average rate of growth (or shrinking) of small perturbations along the solutions of a dynamical system \cite{Lya,Ose68,Ben80}. In regular, nonchaotic systems, the distance between a given trajectory and another one, built from a small perturbation in the initial conditions, remains close to zero, or increases at most algebraically as time evolves. In chaotic systems this distance diverges exponentially in time. The Lyapunov exponent characterizes this instability along the path. When it is positive, the trajectory is extremely sensitive to the initial condition and is called chaotic. Details of the calculation can be found in the Appendix A of Ref. \cite{Cha16}. In Fig. \ref{fig:1}  the dark blue areas are regular, while the light green ones are chaotic. It is important to emphasize that this map of chaos and regularity is a very 
useful guide for the study of the correspondence between the classical and the quantum system. Also, we expect it will be helpful to explore the equilibration and thermalization of initial states in the Dicke model, having it as a reference of 
chaotic behavior for the quantum model.  

\section{Participation ratio as a quantum measure of chaos}

In the context of the kicked top, in \cite{Haake01} F. Haake proposed that the minimum number of eigenstates of the Floquet operator necessary to reconstruct a coherent state, $D_{min}$, could be used to identify chaotic and regular regimes. It was also shown that the $D_{min}$ scales with the size, $L$, of the system, as $\sqrt{L}$ in regular regions and as $L$ in chaotic ones. The proposal depends on the idea of localization. In the regular regions, the coherent states tend to be localized around few states, in correspondence with the classical regular motion, while a coherent state associated with a chaotic region would be delocalized. Therefore, in the thermodynamical limit $L\rightarrow\infty$, an infinitely small fraction of eigenstates ($\sqrt{L}/L=1/\sqrt{L}$) would be enough to reconstruct a coherent state associated with regular regions, whereas, this fraction would go  to a finite value in the chaotic region. It was also suggested that this measure is an analogue of the classical Lyapunov exponent. 
Inspired by this idea we employ the participation ratio of a coherent state spanned in the Hamiltonian eigenstate basis. Each coherent state correlates with one point in the classical phase space of the classical Dicke Hamiltonian, and in our context the number of atoms $j=\mathcal{N}/2$ is the relevant dimension of the system, directly associated with the eigenstate basis. It provides a measure of localization of the quantum state.

The participation ratio $P_{R}$ was introduced several years ago in the context of solid state physics. It was employed to obtain some indication of the number of atoms participating in a normal mode of vibration  \cite{Bell70}, to understand the arguments that have led people to believe that noninteracting electrons in an infinite random system may be either in extended states or in exponentially localized states \cite{Thou74}, and, to describe, for instance,  states of the free electrons in the lowest Landau level \cite{Weg83} and its localization in the presence of a random potential \cite{Hik86}, the  properties of the localized phase of disordered single-particle systems \cite{Zirn86}, delocalization in random banded matrices with strongly fluctuating diagonal elements\cite{Fyo95}, the crossover from Poisson to Wigner-Dyson in many-body Fermi systems \cite{Geo97}, the localization in Fock space in the many-particle excitation statistics of interacting electrons in a finite two dimensional quantum dot \cite{Ber98}, the structure of compound states in the chaotic spectrum of the Ce atom \cite{Fla94} and the entropy production and wave packet dynamics in the Fock space of closed chaotic many-body systems \cite{Fla01}. For the Dicke model it has been employed in the study of the relaxation process and the transition from integrability to chaos \cite{IGM15}.  Also, it has been used to show that the equilibration process depends on the spreading of the initial state over the perturbed basis\cite{Engel15}.

The main advantage over the $D_{min}$ criterion of Haake mentioned above, is that the $P_{R}$ does not requieres a cutoff (the smallest relevant contribution), which could be arbitrary. The $P_{R}$ as a measure of localization has its own scale because it is normalized. That this localization in the space of eigenstates effectively takes place in the Dicke model can be seen in the distribution of the coherent state over the Hamiltonian eigenbasis, as shown in Fig. 3 and 4 of Ref. \cite{Basta16PRE}

For a pure quantum state $|\Psi\rangle$,  expanded in a $N$-dimensional basis $\{|\phi_{k}\rangle\}$, the participation ratio is 
\begin{equation}
P_{R}=\frac{1}{\sum_{k=1}^{N}|\langle \phi_{k}|\Psi\rangle|^{4}}.
\end{equation}
The participation ratio is defined in the interval $P_{R}\in[1,N]$. When $P_{R}=1$ it means the state $|\Psi\rangle$ is identical to one of the states of the basis, and it is considered as having maximum localization. On the other hand, if every state of the basis equally contribute to the state, we would have $|\langle \phi_{k}|\Psi\rangle|=1/\sqrt{N}$. In this case $P_{R}=N$. So, the maximum value of the $P_{R}$ is related to maximum delocalization in the given Hilbert space basis. 

\subsection{The participation ratio in the integrable LMG model}

\begin{figure}
\hspace{100pt}(a) \hspace{150pt}(b)\\
\includegraphics[width=0.45\textwidth]{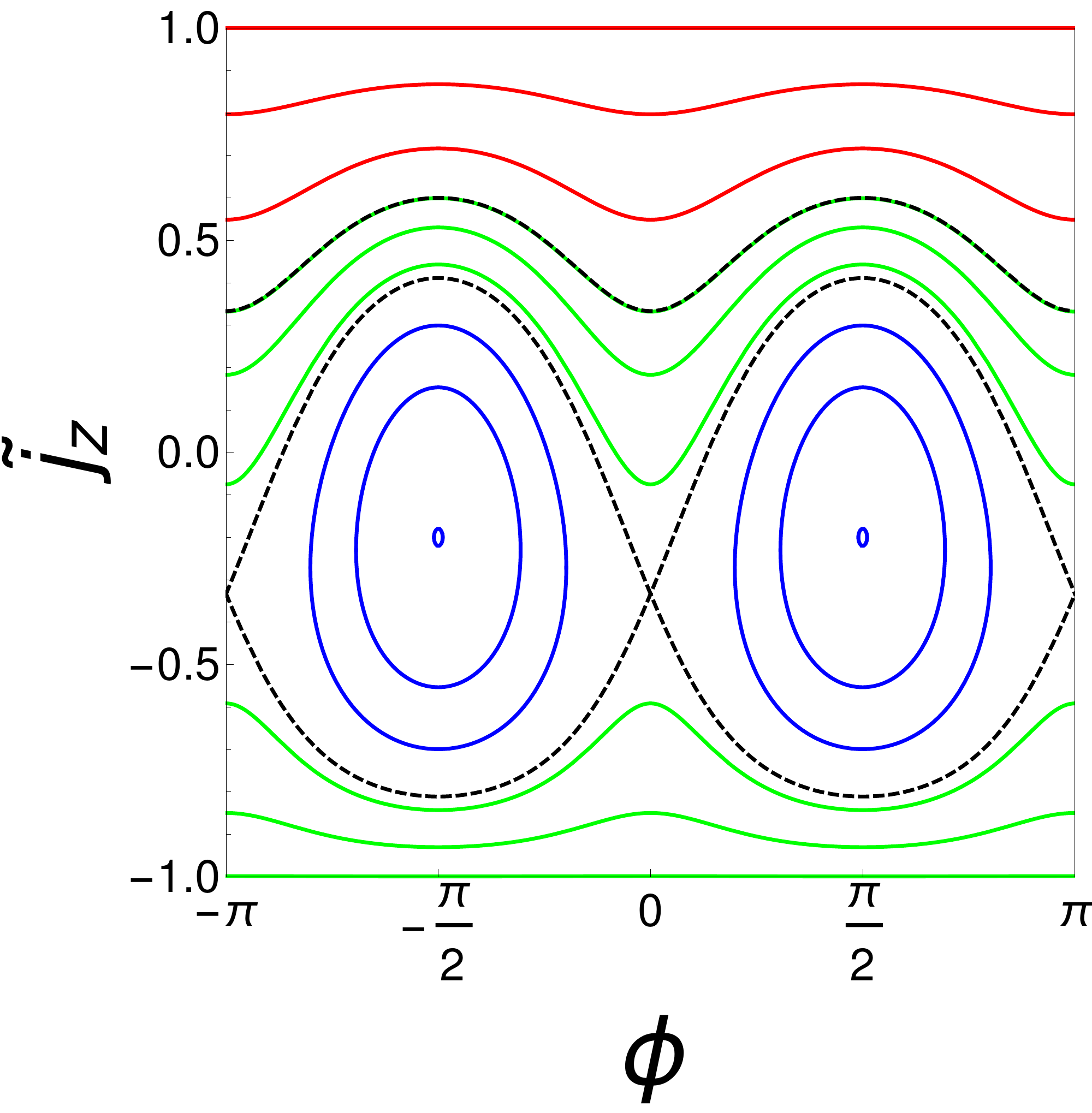}\includegraphics[width=0.52\textwidth]{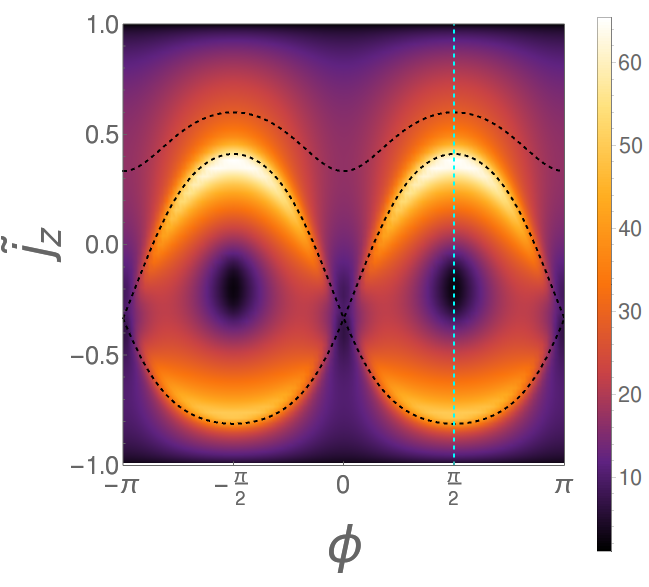}\\
\caption{(a) Constant energy curves for the expectation value of the Lipkin-Meshkov-Glick Hamiltonian in coherent states,   $\langle z | H_L | z\rangle$, for the case  $\gamma_x=-3$, $\gamma_y=-5$, using the canonical coordinates $j_z/J=\tilde{j}_z$ and $\phi$. The degenerate ground state [with energy $E_{min}=-J/2 (\gamma_y + \gamma_y^{-1})$] correspond to the two fixed points located at  the center of the closed orbits (blue lines). The critical energy   $E_{cr}=-J/2 (\gamma_x + \gamma_x^{-1})$ is associated to the separatrix (dashed line) separating the closed orbits from the rotational-type ones (green). At energies above $E=-J$, the lower rotational-type orbits disappear and only the upper ones (red) remain. (b) Participation ratio of  coherent states over the Hamiltonian eigenbasis for the same parameters as panel (a) and for $J=100$.  The participation ratio is  minimal ($\approx 2$) in the ground-state points, and attains its maximal value in  regions associated to the separatrix. The vertical line, $\phi=\pi/2$, indicates the cases studied in  Fig.\ref{fig:3}.}
\label{fig:2}
\end{figure} 

\begin{figure}
\hspace{120pt}(a) \hspace{160pt}(b)\\
\includegraphics[width=0.6\textwidth]{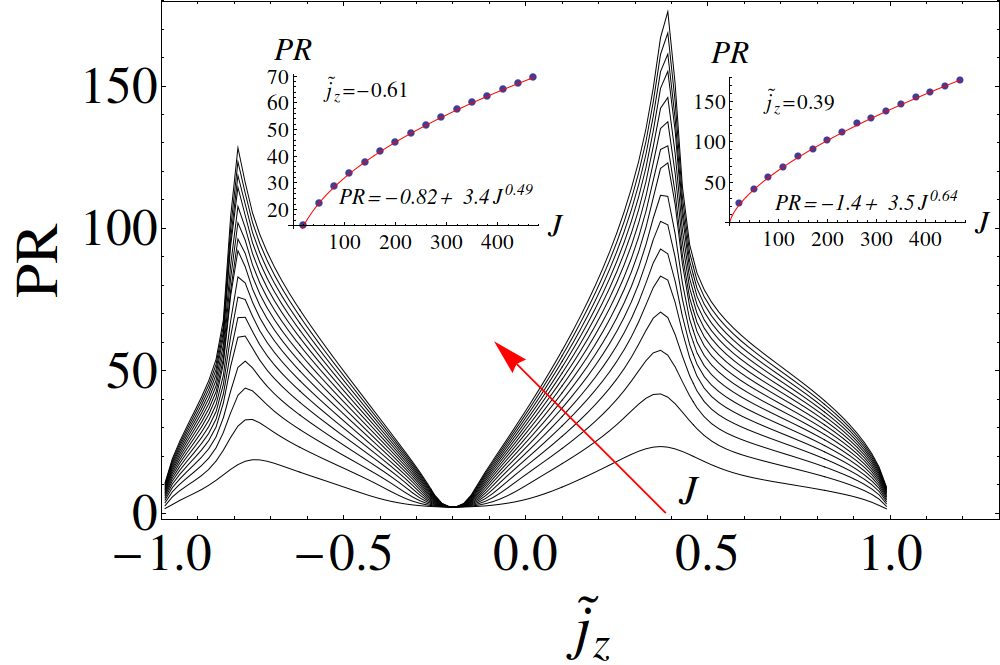}
\raisebox{.3\height}{\includegraphics[width=0.4\textwidth]{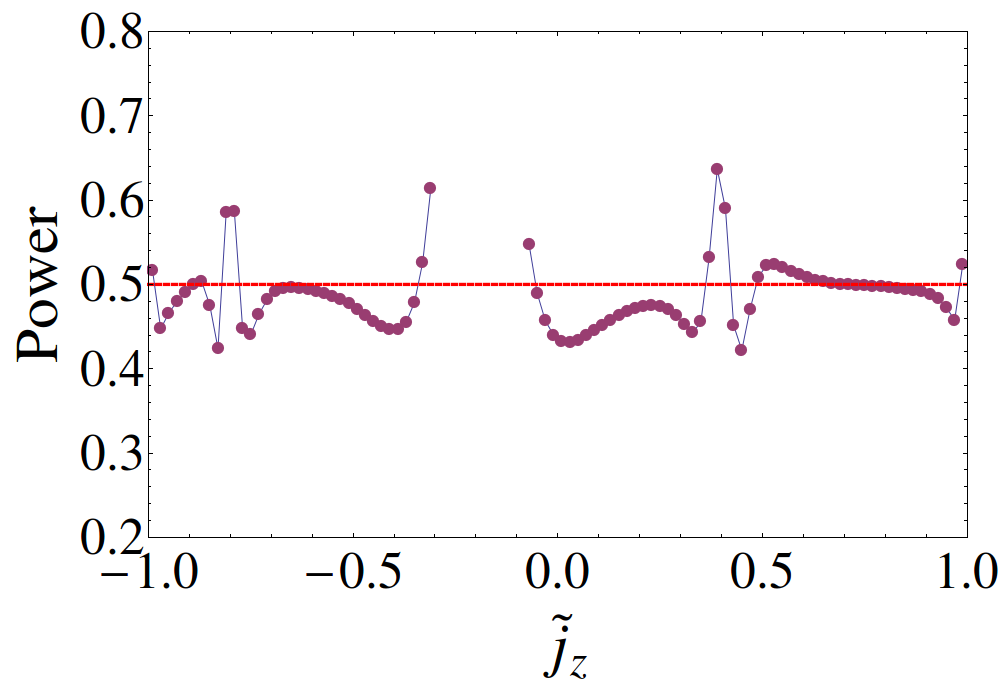}}
\caption{  (a) Participation ratio for the line $\phi=\pi/2$,  for  $J=20,50,80,...,470$  (red arrow signals the increase of $J$). Except in the ground-state energy region ($\tilde{j}_z\sim -0.2$), where  $P_R\approx 2$ for all $J$, the participation ratio increase with $J$. The increase is always slower than $J$. In the insets the scaling of the participation ratio is shown for two particular cases, one corresponding to the separatrix where the $P_R$ takes its maximal value. The numerical values are fitted with functions of the form $P_R= a+ b J^c$ (the values of the fitted parameters are indicated in the insets). In panel (b) the   power of $J$ in the fitted functions (parameter $c$ in $P_R= a+ b J^c$) is shown for the coherent states along the line $\phi=\pi/2$, excluding the ground-state energy region, where $P_R\approx 2$. Observe that the powers are smaller than one and very close to $1/2$ (horizontal dashed line), which suggest that in the limit $J\rightarrow\infty$, the participation ratio divided by $J$ goes to zero.}
\label{fig:3}
\end{figure} 

As a simple  example to illustrate the behavior of the participation ratio in a finite Hilbert space, we choose the Lipkin-Meshkov-Glick (LMG) model, parametrized as follows \cite{OCasta06}
$$
H_L= J_z+ \frac{\gamma_x}{2J-1} J_x^2+ \frac{\gamma_y}{2J-1} J_y^2,
$$
and calculate the participation ratio of Bloch Coherent states, $|z\rangle$, over the Hamiltonian eigenbasis for the set of parameters $\gamma_x=-3$, $\gamma_y=-5$. Some  energy curves of $\langle z | H_L | z\rangle$ are illustrated in panel (a) of Fig.\ref{fig:2} using the canonical variables $\tilde{j}_z$ and $\phi$ defined above. 

Since the LMG model has only  one degree of freedom, these curves are equal to the trajectories of the associated classical model. We can see that its phase space  is very similar to that of the simple pendulum. At $\phi=\pm \pi/2$ and $\tilde{j}_z\approx -0,2$ it has two stable fixed points, which correspond to the  minimal energy given by $E_{min}=-J/2 (\gamma_y + \gamma_y^{-1})$. For slightly larger energies, there are  librational trajectories around the elliptic fixed points. At the critical energy  $E_{cr}=-J/2 (\gamma_x + \gamma_x^{-1})$, there exists a separatrix associated to  hyperbolic fixed points located at  $\phi=0$ and $\pi$. For larger energies than $E_{cr}$, the orbits become of rotational type. Every orbit in the upper part of the phase space has an associated orbit with the same energy located in the lower part. The existence of degenerate orbits occurs until the energy $E=-J$, where the lower orbits disappear, and only the upper orbits remain until the maximal energy available $E=+J$. 

In Fig.\ref{fig:2}(b), a density plot of the participation ratio of coherent states over the Hamiltonian eigenbasis is shown for the same parameters $\gamma_x$ and $\gamma_y$ and with $J=100$.  The participation ratio attains its maximal values in regions associated with the separatrix at critical energy  $E_{cr}=-J/2 (\gamma_x + \gamma_x^{-1})$. This property  is used in Ref.\cite{Engel15} to propose an experimental protocol to detect the critical behaviour of the quantum Density of States (the so-called Excited-State Quantum Phase Transition) which takes place at this critical energy.  

As mentioned above,
the LMG model has only one degree of freedom and is necessarily integrable. Consequently, even if in the region of the separatrix  the participation ratio takes larger values, it is expected that the participation ratio scales slower than $J$ for every coherent state.
This means that, in the thermodynamic limit, $J \rightarrow \infty$, any coherent state becomes localized in the eigenstate basis.
This expectation is explored  in Fig.\ref{fig:3}, where the dependence of the participation ratio on $J$ is shown for coherent states along the line $\phi=\pi/2$. In panel (a) the participation ratio is shown for $J=20,50,80,...,440,470$ as a function of variable $\tilde{j}_z=j_z/J$. 
 
In the ground-state region the participation ratio takes a value close to $2$ independently of $J$. This result is not surprising because for the chosen parameters,  the ground ($|GS\rangle$) and first excited ($|1ES\rangle$) states are almost degenerate  and approximately  given by a linear combination of two coherent states \cite{OCasta06} 
$$|GS\rangle\approx\frac{1}{\sqrt{2}}\left( |z(j_{zo},\phi=\pi/2)\rangle + |z(j_{zo},\phi=-\pi/2)\rangle\right) 
$$
 and 
  $$|1ES\rangle\approx\frac{1}{\sqrt{2}}\left( |z(j_{zo},\phi=\pi/2)\rangle - |z(j_{zo},\phi=-\pi/2)\rangle\right),$$
where $z(j_{zo},\phi=\pm\pi/2)$ 
are the coherent parameters associated to the degenerate minima of the energy surface shown in  Fig.\ref{fig:2}(a). Therefore, the coherent states close to the ground-state energy are $|z(j_{zo},\phi\approx\pi/2)\rangle\approx \frac{1}{\sqrt{2}}(|GS\rangle+|1ES\rangle)$, with  participation ratio  close to $2$. 

In the rest of the interval along the line $\phi=\pi/2$, the $P_{R}$ increases clearly with $J$, as can be seen in Fig.\ref{fig:3}(a). In the insets, the dependence of the participation ratio  with $J$ is shown for two representative cases. One of them corresponds to  the case where the participation ratio attains its maxima.  In both cases, a fit of the dependence on $J$ is made using functions of the form $P_R=a+bJ^c$. 
In both cases the power of $J$ is smaller than one and close to $1/2$. Similar fittings are made for points along the line $\phi=\pi/2$ (excluding the ground-state energy region) and the resulting powers of $J$ are shown in Fig.\ref{fig:3}(b). It is clear that the growing of the participation ratio is approximately  given by a square root dependence, $J^{1/2}$. Consequently, in  the limit $J\rightarrow\infty$,  the ratio $P_R/J$  is expected to go to zero. It is worth mentioning that also for the ground state energy region, where $P_R\approx 2$,  the ratio $P_R/J$ goes to  zero  in the limit $J\rightarrow\infty$.   

The behavior of the participation ratio in the integrable LMG model has to be contrasted with the results reported for the kicked top \cite{Haake01}, which has the same Hilbert space as the LMG model. As mentioned before, in the kicked top,  the number of eigenstates (of the Floquet operator in this case) that suffices to build a coherent state associated to a classical chaotic region ($D_{min}$) depends linearly on $J$; consequently the limit $J\rightarrow \infty$ gives a non-zero ratio $D_{min}/J$.

\subsection{The participation ratio in the non-integrable Dicke model}

Now we turn our attention to the two-degrees of freedom Dicke model. In this case the    
$P_{R}$ of the coherent state $|\alpha_{0},z_{0}\rangle$, which identifies a point in the classical phase-space, expanded in the Hamiltonian eigenstates $|E_k\rangle$,  is 
\begin{eqnarray}
P_{R}&=\frac{1}{\sum_{k}|\langle E_{k}|\alpha_{0},z_{0}\rangle|^{4}}=\frac{1}{\sum_{k}Q_{k}^{2}(\alpha_{0},z_{0})},
\end{eqnarray}
where $Q_{k}^{2}(\alpha_{0},z_{0})$ is the Husimi function of the $k-th$ eigenstate $|E_{k}\rangle$ of the Dicke Hamiltonian. The Husimi or $Q$ function, $Q_{k}(\alpha,z)=|\langle z, \alpha |E_{k}\rangle|^2$, is one of the simplest distributions of quasiprobability in phase space. It has a well-defined classical limit and allows to study the classical-quantum correspondence \cite{Bak13}. Also, in the thermodynamical limit it reduces to a classical probability function on phase space obeying the Liouville equation \cite{Alt12}. The Husimi function has been employed in the Dicke model by several authors to study the quantum-classical transition and equilibration \cite{Alt12,Bak13}, the wave functions of individual states  \cite{MAM91,Bak13}, and the ground-state QPT \cite{Rom12,Real13}.

In order to better compare the quantum results (based on the $P_R$) with the classical ones (based on the Lyapunov exponent), we define binary quantities. We consider the quantity 
 $P_{R}/\mathcal{N}$, and  we assign $P_{Rbin}=0$ when $P_{R}/\mathcal{N}<1$ and $P_{Rbin}=1$ when $P_{R}/\mathcal{N}>1$.
This binary quantity gives  a very simple criterion allowing to distinguish roughly when the participation ratio scales slower than the number of atoms. Analogously, for the Lyapunov exponent we define $\Lambda_{bin}$ as $\Lambda_{bin}=0$ if $\Lambda<0.004$ and   $\Lambda_{bin}=1$ if $\Lambda>0.004$. The numerical  tolerance $\Lambda_T=0.004$ employed to distinguish a chaotic from a regular classical trajectory,  is rooted in the numerical precision employed in solving the equations of motion and has been determined by comparing the obtained  Lyapunov values with the Smaller Alignment Index (SALI) method \cite{Skokos01}, which allows to determine reliably if a given set of initial conditions corresponds to a regular or chaotic dynamics. Nevertheless, care has to be taken when the respective values, $P_{R}/\mathcal{N}$ and $\Lambda$, are close to the tolerance limits employed ($1$ and $0.004$ respectively).  
In the following we are going to explore these quantities along the energy surfaces. 

\subsection{The efficient coherent basis} \label{ECB}

\begin{figure}
\begin{tabular}{cc}
(a) &(b)\\
\includegraphics[width=0.5\textwidth]{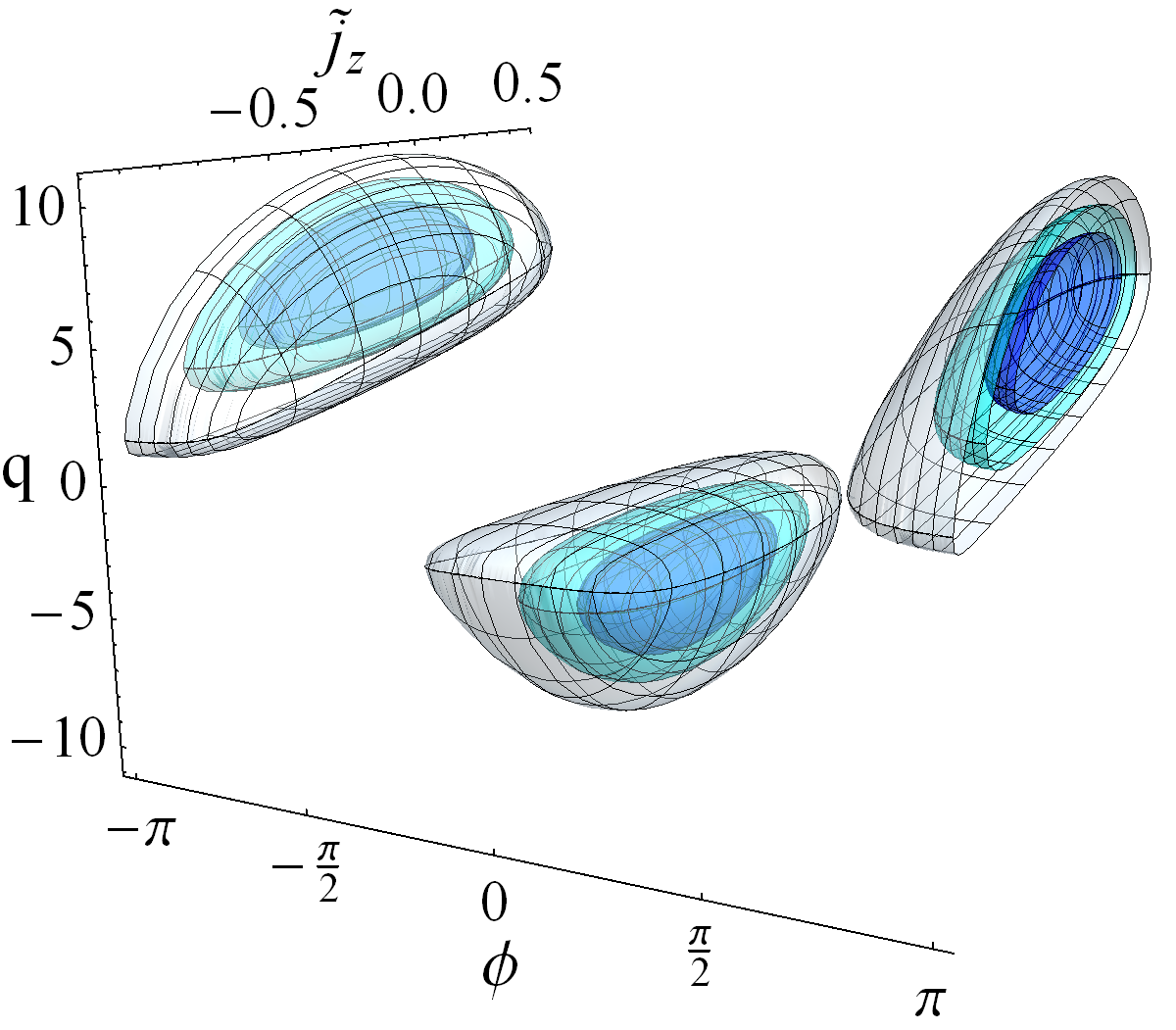}&\includegraphics[width=0.5\textwidth]{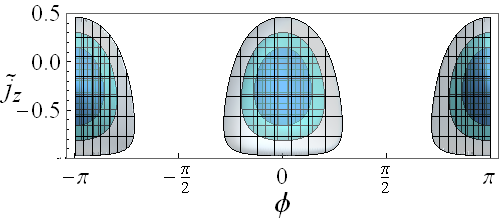}\\
(c) &(d)\\
\includegraphics[width=0.5\textwidth]{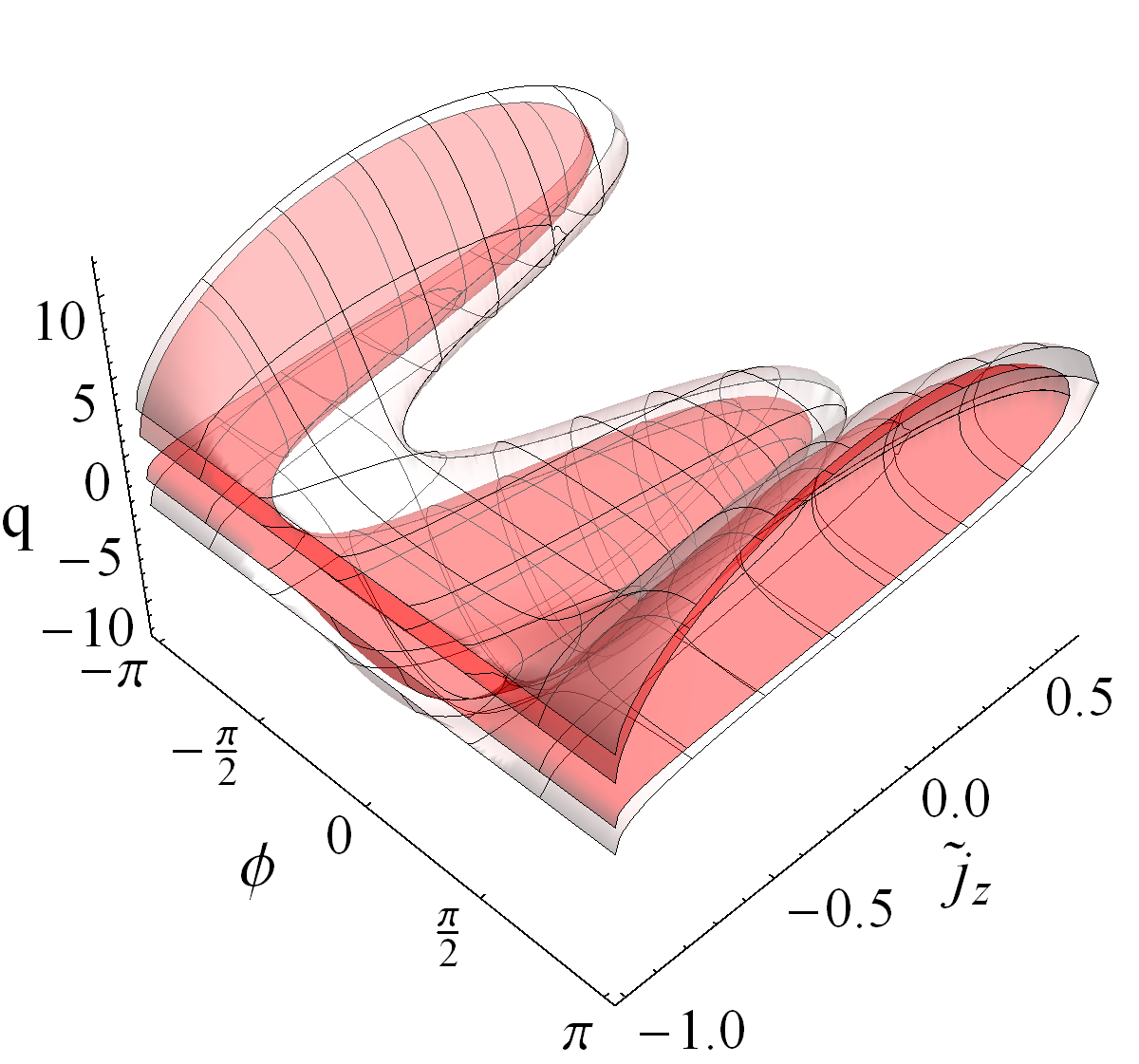}&\includegraphics[width=0.4\textwidth]{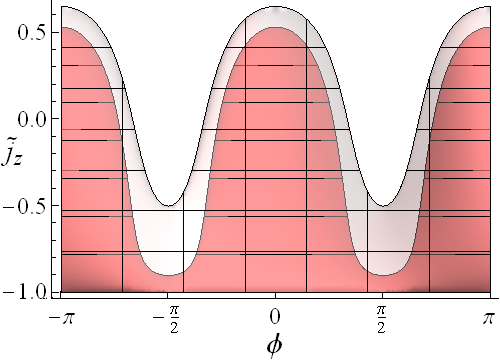}
\end{tabular}
\caption{ Poincar\'e surfaces ($p=0$) -(a, c)-, and their projection in the $(\tilde{j_{z}},\phi$) plane -(b, d)-, for the  energies indicated in Fig.\ref{fig:1}. 
(a, b) show the surfaces for  energies below the ESQPT, from the inner to the outer surface these energies are, respectively, $\epsilon/\omega_o=-1.8,-1.5$ and $-1.1$. (c, d) is similar for  energies above the ESQPT, $\epsilon/\omega_o=-0.9$ and $-0.5$ for the inner and outer surface respectively.  Every surface is formed by an upper and lower shell, $q_+$ and $q_-$, given by the solutions shown in  Eq.(\ref{qpm}).}
\label{fig:4}
\end{figure}

In the evaluation of the participation ratio $P_R$ the challenge is to obtain the exact quantum eigenstates $|E_{k}\rangle$ of the Dicke Hamiltonian. Being the Hamiltonian non-integrable, the solutions (to obtain the  spectrum and the eigenfunctions) must be numerical. On the other hand, given that the Hilbert space is infinite, the numerical calculations must be performed in a truncated space, and the convergence of each one of the eigenstates which are relevant in the expansion of a coherent state must be guaranteed.  To perform this task we employ the efficient coherent basis (ECB), which is the exact Hamiltonian eigenbasis in the limit $\omega_0=0$ \cite{Chen0809,Basta11}. 

The ECB is built from vacuum sates, $|0\rangle_{m_x}$, defining new bosonic displaced operators $A=a+\frac{2\gamma}{\omega\sqrt{\mathcal{N}}}\,J_{x}$, 
 \begin{equation}
|N;j,m_x\rangle=\frac{(A^\dagger)^N}{\sqrt{N!}}|0\rangle_{m_x}. 
 \end{equation}
These vacuum states are obtained from rotated (by $-\frac{\pi}{2}$ around the $y$-axis) atomic states,
 eigenstates $|j,m_x\rangle$ of $J_{x}$
 $$
 |0\rangle_{m_x}=|\alpha_{m_x}\rangle|j, m_x\rangle,
 $$
 where $|\alpha_{m_x}\rangle$ is a boson coherent state with Glauber parameter $\alpha_{m_x} =-2\gamma m_x/(\omega\sqrt{\mathcal{N}})$, which guarantees that $A|0\rangle_{m_x}=0$.

The benefits of employing this particular basis  are notorious. It allows to obtain a number of converged eigenstates orders of magnitude larger than that which can be obtained with the same cut-off in the standard Fock basis, based in the number of photons and the eigenstates of $J_z$. It is particularly useful in the study of large systems in the superradiant phase, which rapidly become numerically intractable in the Fock basis \cite{Basta11,Basta12,Basta14}. Details and explicit expressions for the evaluation of the Husimi function, and for the participation ratio in the efficient coherent basis are given in Appendix C of Ref. \cite{Basta16PRE}.

Finally, at this point it is important to emphasize that, even though, the Hilbert space of the Dicke model is infinite due to the presence of the bosonic operators, the wave functions have a gaussian profile and thanks to this a cutoff is enough to have a complete wave function, as it is shown in Ref. \cite{Basta14}. Then, even though there is no absolute value for the participation ratio of a delocalized wave function like in the case of the LMG model, given a cut-off $N_{max}$, a maximum delocalized state would have a participation ratio of $(2j+1)(N_{max}+1)$.

\section{Participation ratio over different energy sections}

We present results for different energies with a single value of the coupling. They are shown as red dots in Fig. 1. We have selected the superradiant phase with $\gamma=2.0 \, \gamma_{c}$ in resonance $\omega=\omega_{0}=1$, as the correspondent classical Hamiltonian of the Dicke model shows different mixings of chaos and regularity, from a complete region of regularity at low energies, to full ergodic chaos \cite{Cha16}. 

To choose the classical canonical coordinates which define each coherent state, we 
consider classical Poincar\'e surfaces of given energy, fixing $p=0$. We find  the variables $\phi$, $\tilde{j_{z}}$, and $q$ which satisfy $h_{cl}(q,p=0,\tilde{j_{z}},\phi)=\epsilon$.  The choice $p=0$ ensures a broad sampling of  classical orbits because all of them intersects these Poincar\'e energy surfaces. Under these conditions, we have two different values of 
$q$, 
\setlength{\mathindent}{0pt}
\begin{equation}
q_{\pm}(\tilde{j_{z}},\phi,\epsilon)=-\frac{2\gamma}{\omega}\sqrt{1-\tilde{j_{z}}^{2}}\,\cos\phi\,
\pm\sqrt{\frac{4\gamma^{2}}{\omega^{2}} \left(1-\tilde{j_{z}}^{2}\right)\,\cos^{2}\phi+\frac{2}{\omega}\left(\epsilon-\omega_{0}\tilde{j_z}\right)}.
\label{qpm}
\end{equation}
\setlength{\mathindent}{15pt}
For a given energy $\epsilon$, each pair of values of $\tilde{j_{z}}$ and $\phi$ identify unambiguously a single point in the phase space within each of the surfaces $q_{\pm}$. The Poincar\'e surfaces for the selected energies in this study are shown in Fig.\ref{fig:4}. 

\begin{figure}
\centering
\begin{tabular}{rccr}
 & $q_+$ & $q_-$\\
 & \includegraphics[width=0.31\textwidth]{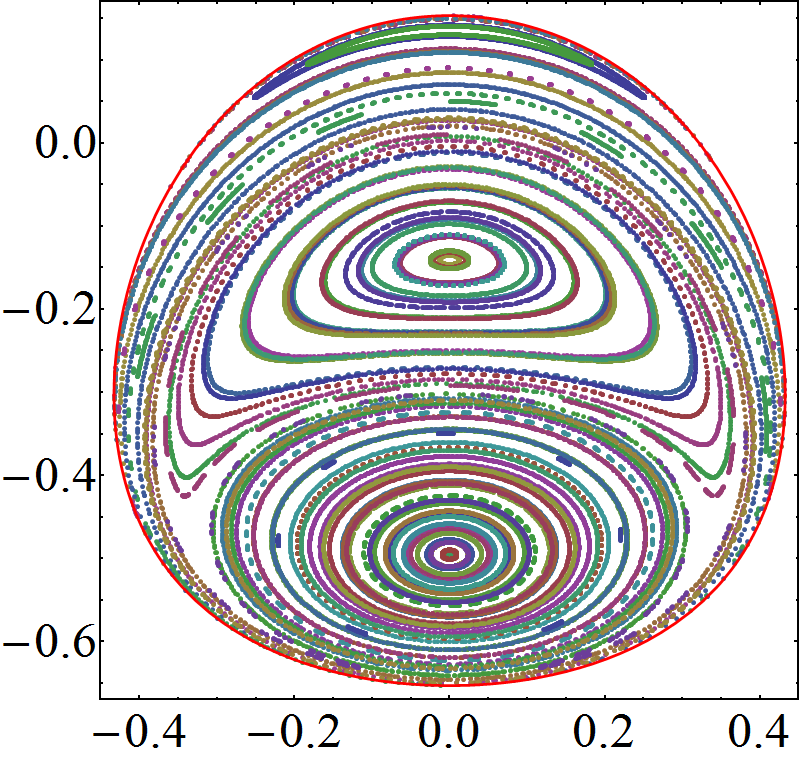}& \includegraphics[width=0.31\textwidth]{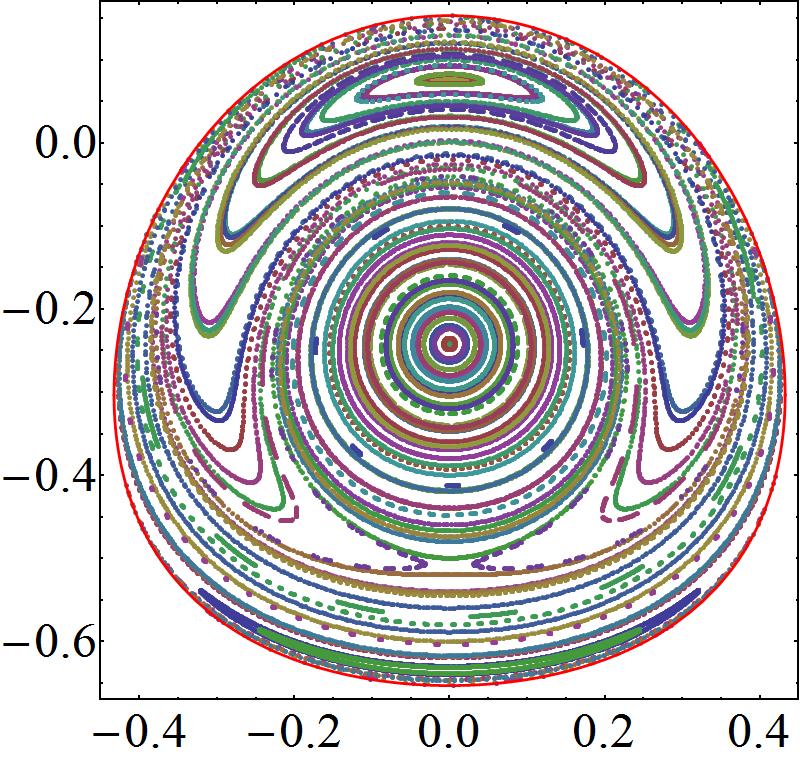}&$\epsilon=-1.8 \omega_o$\\
  & \includegraphics[width=0.31\textwidth]{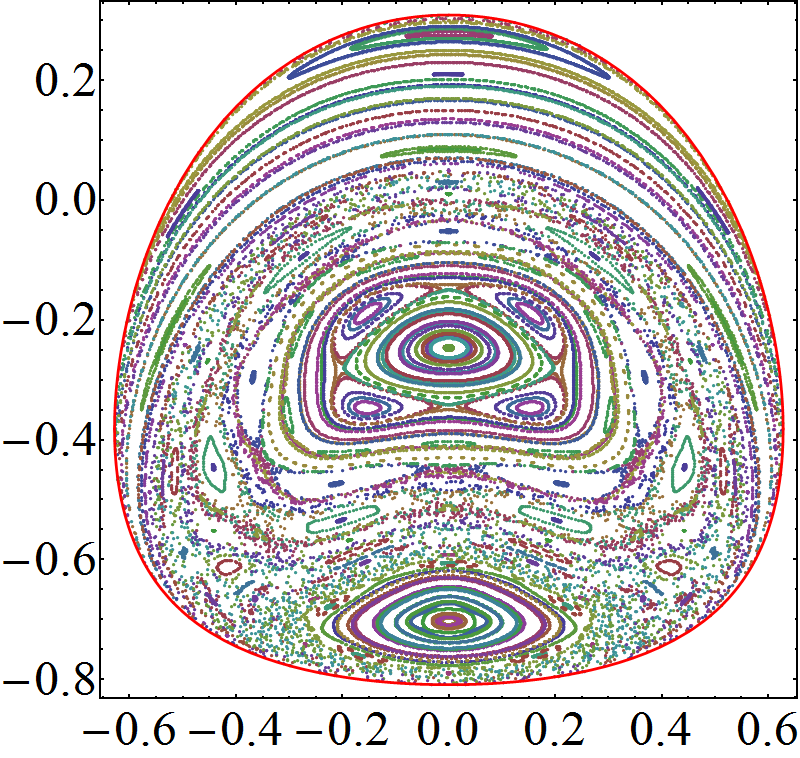}& \includegraphics[width=0.31\textwidth]{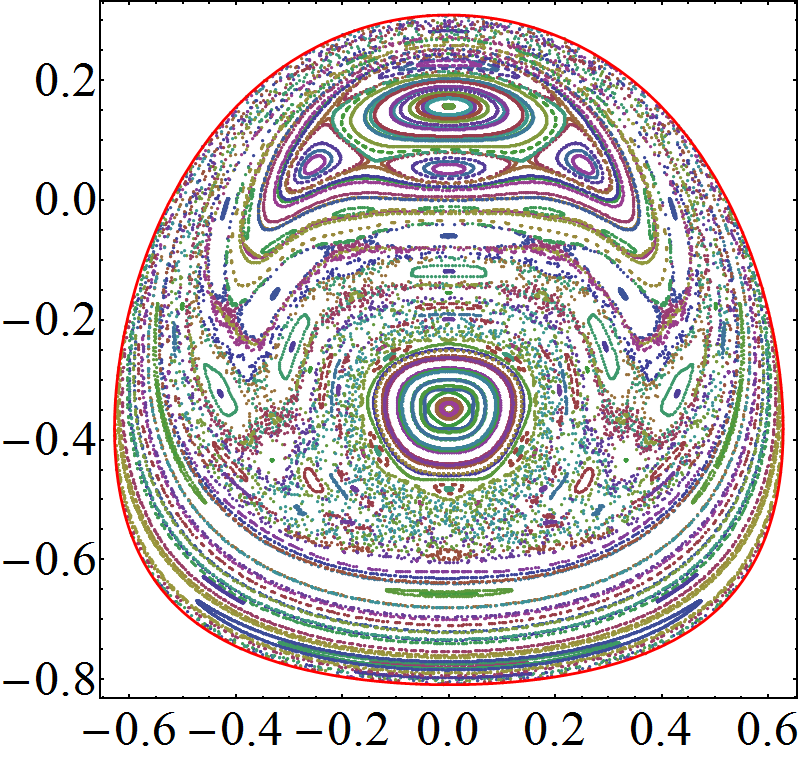}& $\epsilon=-1.5 \omega_o$\\
 $\tilde{j}_z\!\!\!\!\!\!$  & \includegraphics[width=0.31\textwidth]{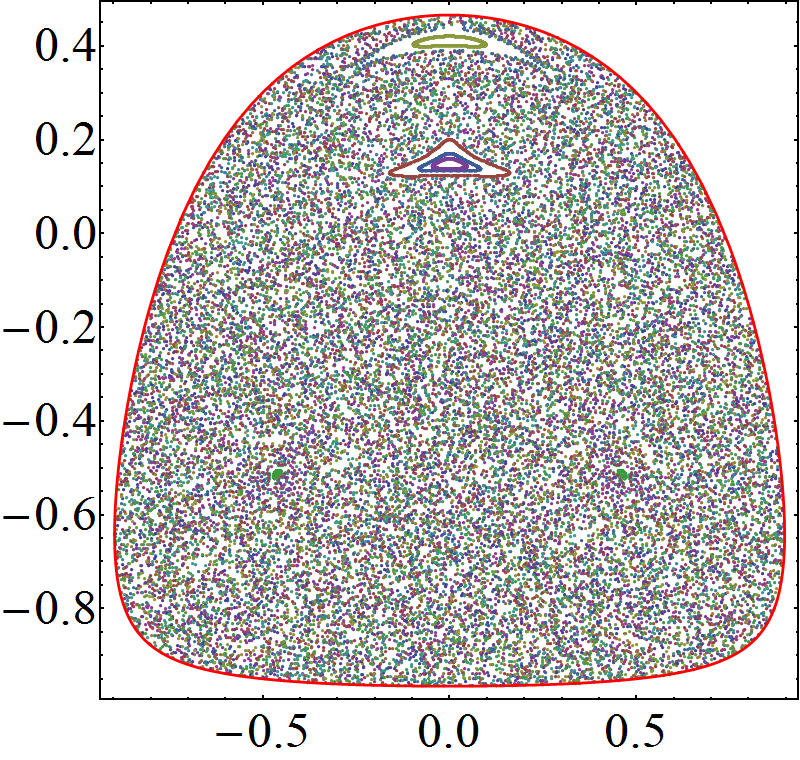}& \includegraphics[width=0.31\textwidth]{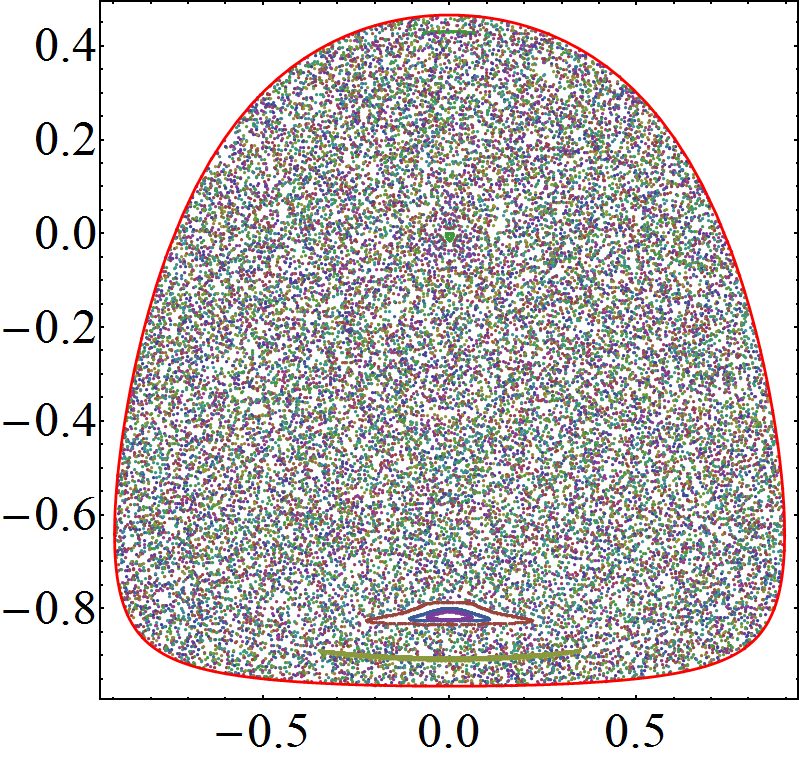}&$\epsilon=-1.1 \omega_o$ \\
    & \includegraphics[width=0.31\textwidth]{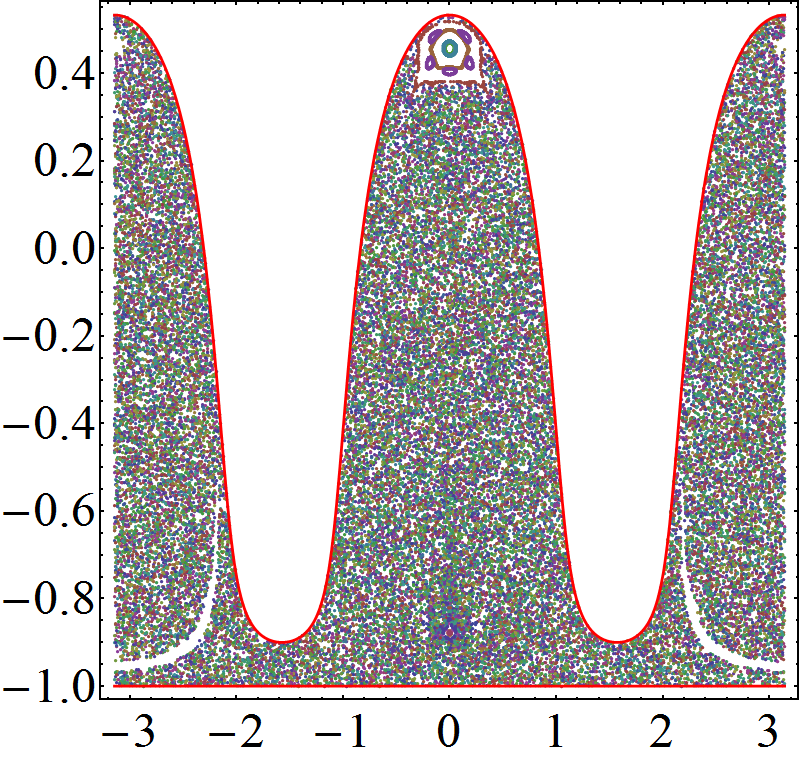}& \includegraphics[width=0.31\textwidth]{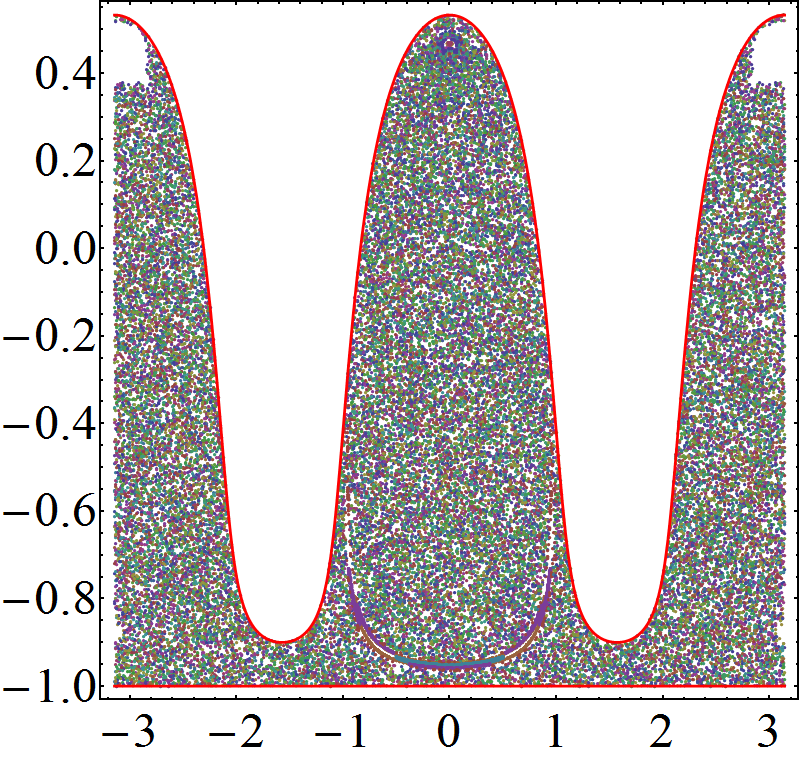}&$\epsilon=-0.9 \omega_o$ \\
     & \includegraphics[width=0.31\textwidth]{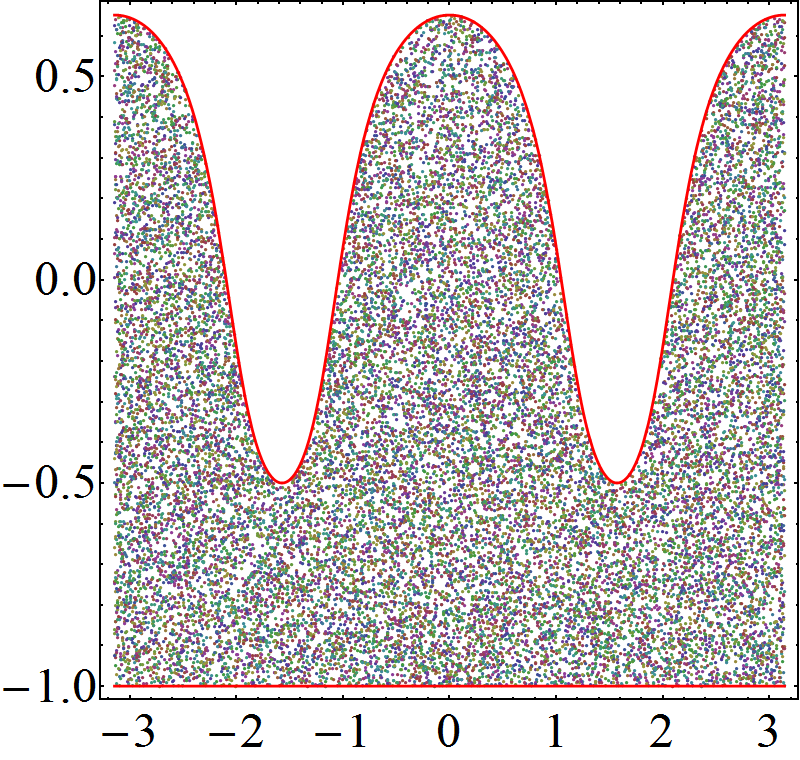}& \includegraphics[width=0.31\textwidth]{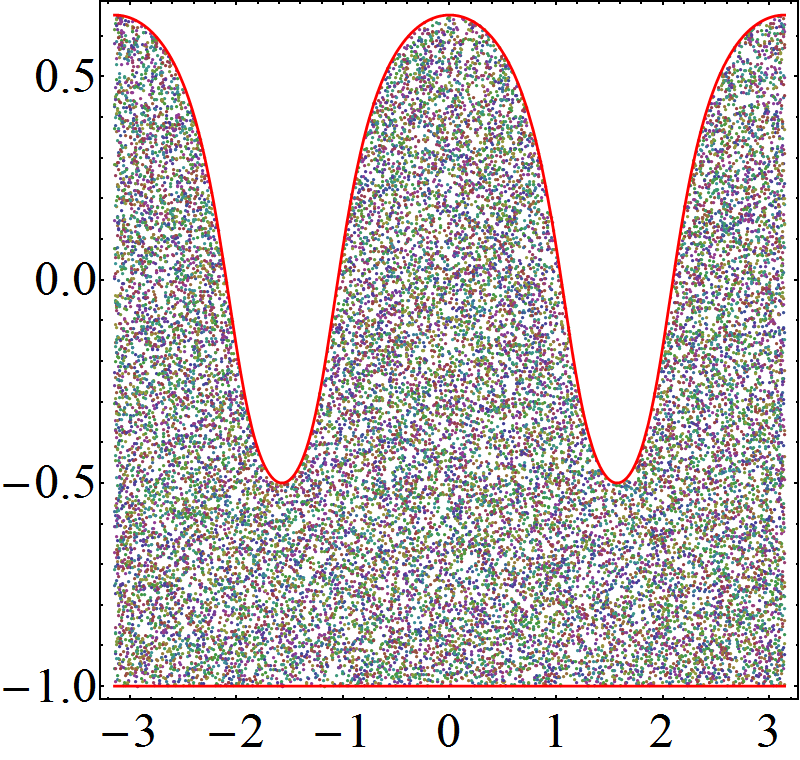}& $\epsilon=-0.5 \omega_o$\\
 &\multicolumn{2}{c}{$\phi$}
\end{tabular}
\caption{Poincar\'e sections ($p=0$) projected over the $\tilde{j}_z$-$\phi$ plane,  for the energies indicated in Fig.\ref{fig:1}.  Left and right  columns correspond, respectively,  to the projections  of  the  $q_+$ and $q_-$ shells of the  Poincar\'e surface.   }
\label{fig:5}
\end{figure}

With these elements we define coherent states $|\alpha,z\rangle$ with $\alpha=\sqrt{\frac{j}{2}}q_{\pm}\left(\epsilon,\tilde{j}_{z},\phi\right)$ and $z=\sqrt{\frac{1+\tilde{j}_{z}}{1-\tilde{j}_{z}}}e^{i\phi}$. The same Poincar\'e surfaces are used  to obtain   Poincar\'e sections of the classical model, which provide a qualitative insight of the presence of chaos \cite{Par89,Stro94,Basta15PS}. Projections of these  Poincar\'e sections in the plane $\tilde{j}_z$-$\phi$,  are  shown in Fig.\ref{fig:5} for the five energies shown with red dots in Fig. \ref{fig:1}, and for both energy surfaces  $q_+$ (left) and $q_-$ (right).

In the following we present a detailed comparison between the $P_{R}$ as a quantum measure of chaos  and the Lyapunov exponent. 
These results complement and expand those presented previously in Figs.  6 and 7 of Ref. \cite{Basta16PRE}, where the connection between the classical Lyapunov exponents and the quantum participation ratio of coherent states on the eigenenergy basis was exhibited for the first time restricted to the energy value $\epsilon=-1.4\, \omega_{0}$.
Due to the numerical effort in calculating the $P_{R}$, we restrict ourselves to the line with $\phi=0$, for each of the  $q=q_{+}$ and $q=q_{-}$ surfaces,    for the   five representative energies: $\epsilon=-1.8 \omega_0$, $\epsilon=-1.5 \omega_0$, $\epsilon=-1.1 \omega_0$, $\epsilon=-0.9 \omega_0$, and $\epsilon=-0.5 \omega_0$.  All the calculations were done fixing the  number of atoms to  $\mathcal{N}=160$ ($j=80$). In the binary case, as both $\Lambda_{bin}$ and $P_{Rbin}$ are zero or one, they are slightly displaced vertically for better visualization.

\subsection{Regular region $\epsilon=-1.8 \,\omega_{0}$}

\begin{figure}
\centering
\begin{tabular}{cc}
(a)&(b)\\
\vspace{-1.3 cm}\\
\includegraphics[width=0.5\textwidth]{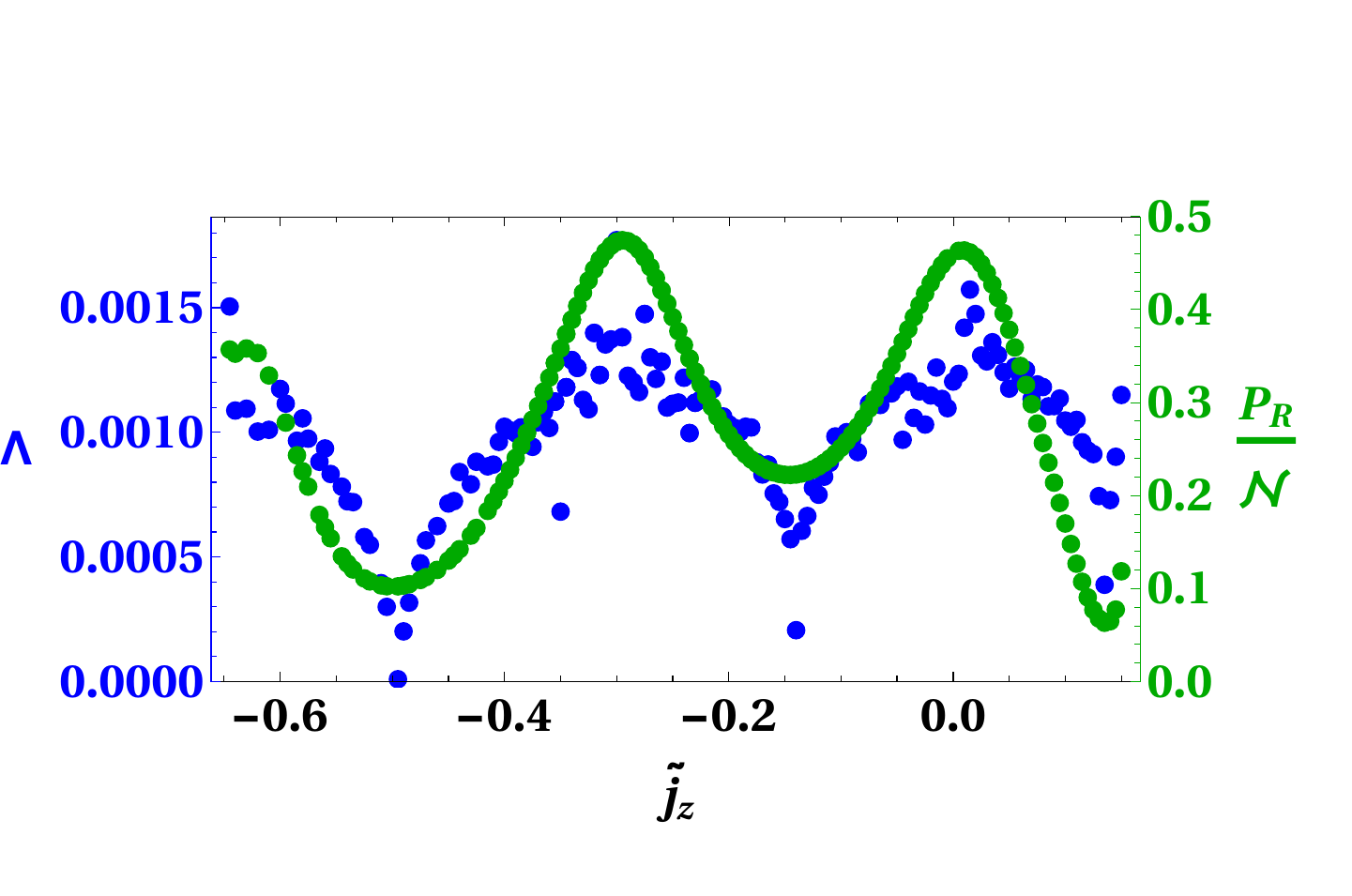} & \includegraphics[width=0.5\textwidth]{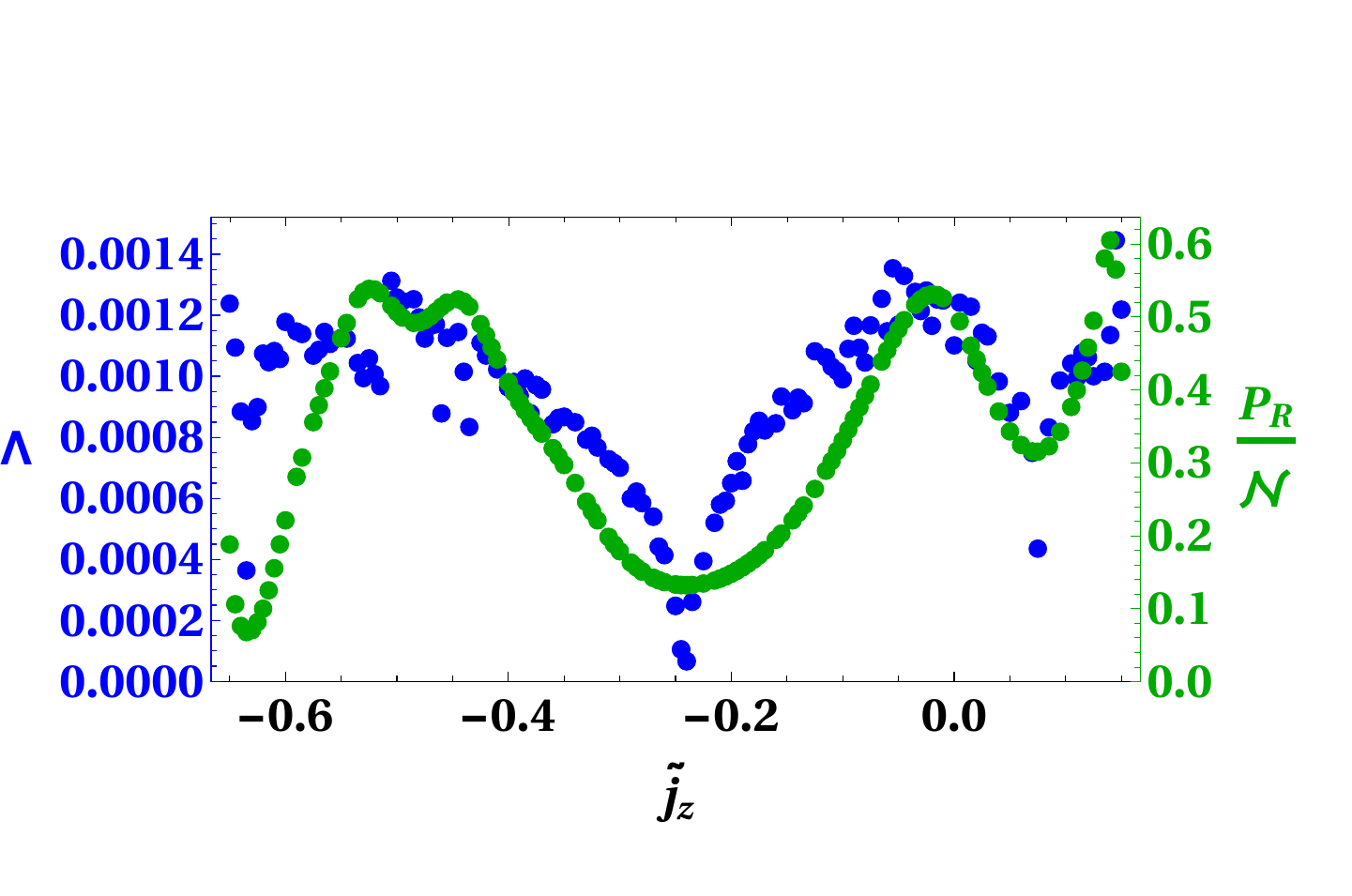} \\
(c)&(d)\\
\vspace{-1.3 cm}\\
\includegraphics[width=0.5\textwidth]{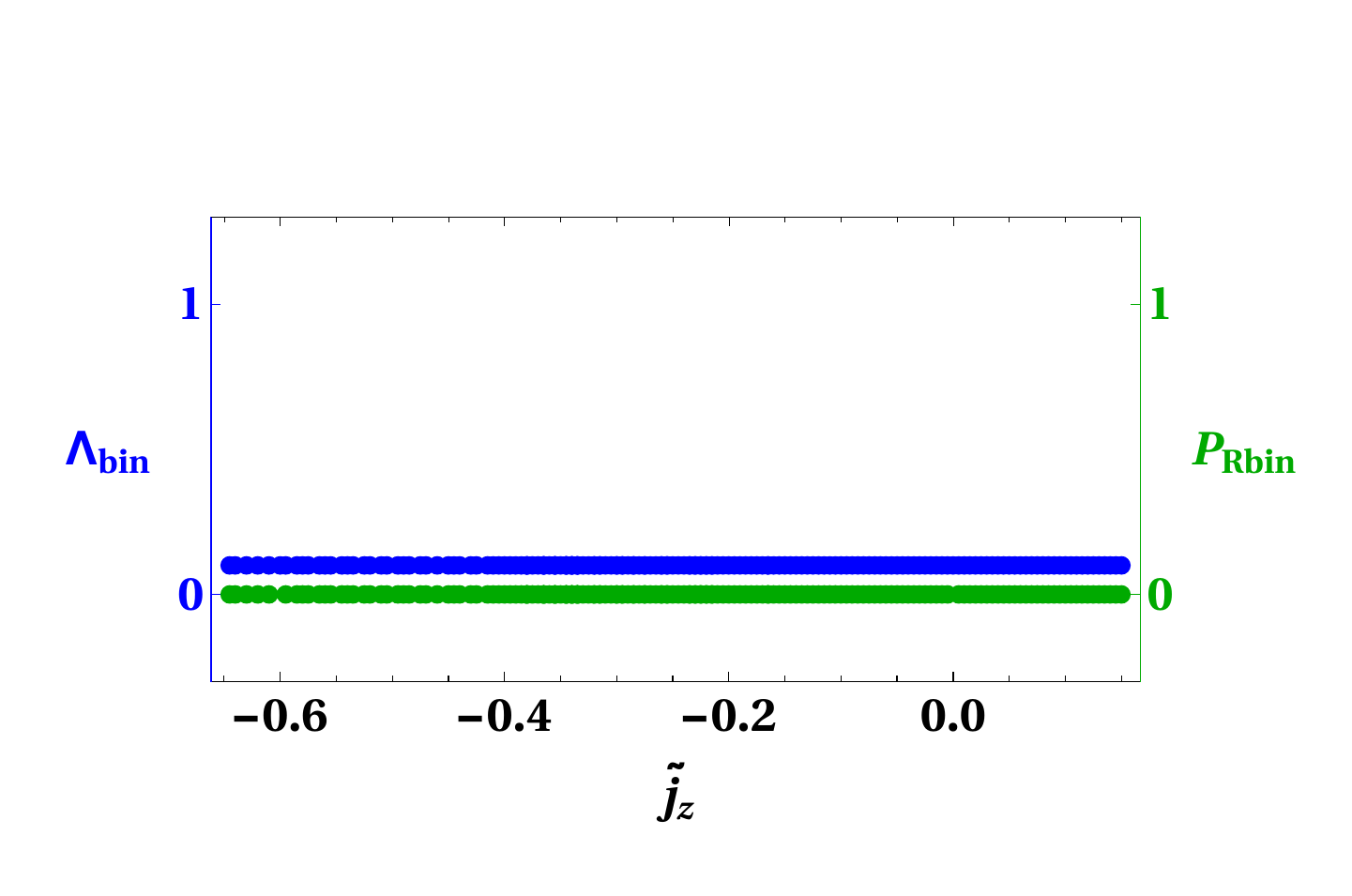} & \includegraphics[width=0.5\textwidth]{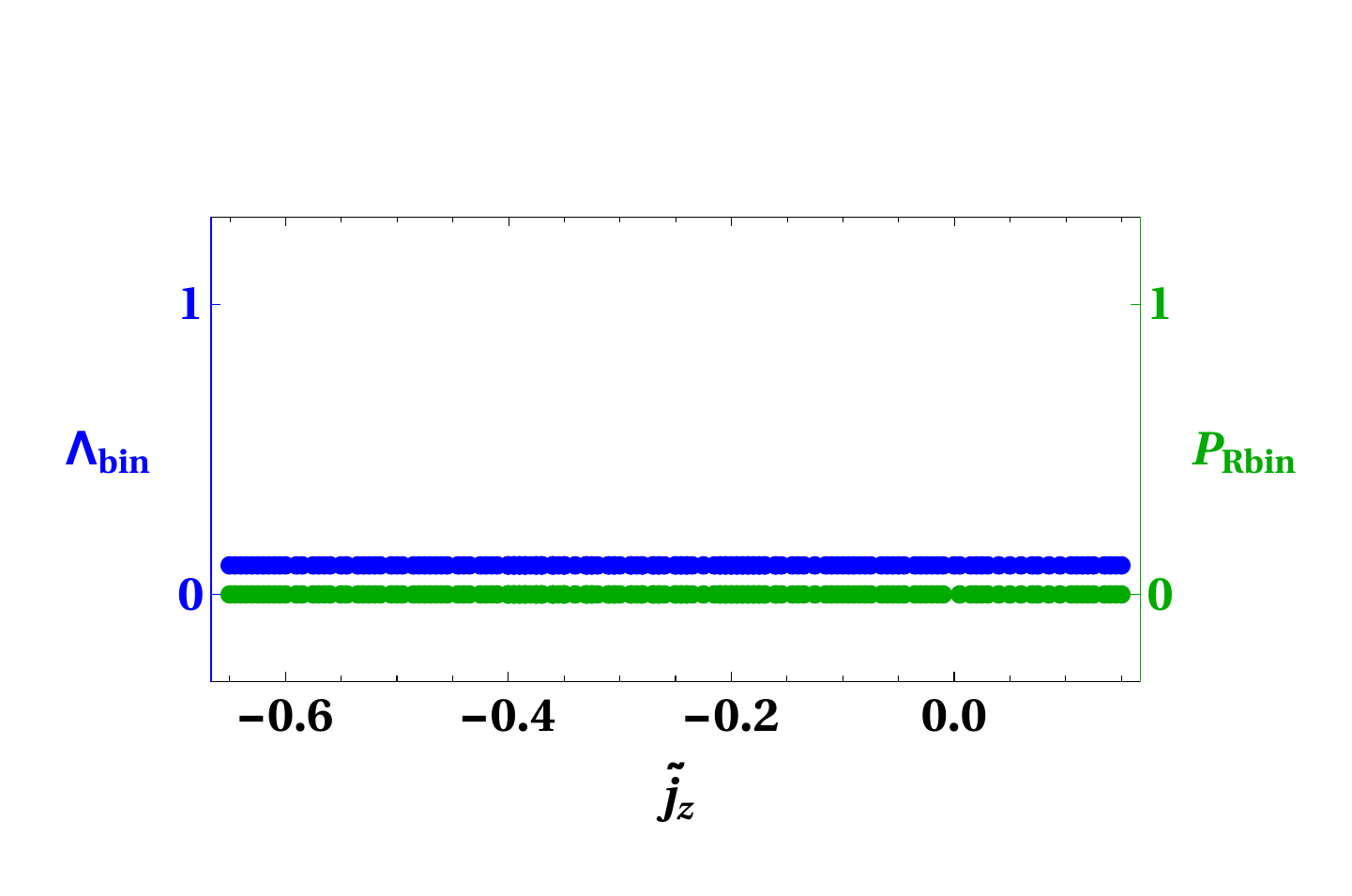} \\ 
\end{tabular} 
\caption{Energy $\epsilon=-1.8\,\omega_{0}$. Participation ratio (green) and Lyapunov exponent (blue) (a, b), and their binary versions (c, d), for the $q_{+}$ surface (a, c) and $q_{-}$ surface (b, d), for the case $\omega=\omega_0$, $\gamma=2\gamma_c$ and $j=80$. According to the classical Poincar\'e surface sections of Fig.\ref{fig:5},  at this energy the dynamics is almost completely regular. A noticeably correlation between the classical Lyapunov coefficient and quantum  $P_{R}$ is observed, even if in both cases the Lyapunov and $P_R/\mathcal{N}$ are below the numerical tolerance used to determine that the dynamics is regular, as can be observed in the panels (c) and (d).  }
\label{fig:6}
\end{figure}
 \begin{figure}[t]
\centering
\includegraphics[width=0.5\textwidth]{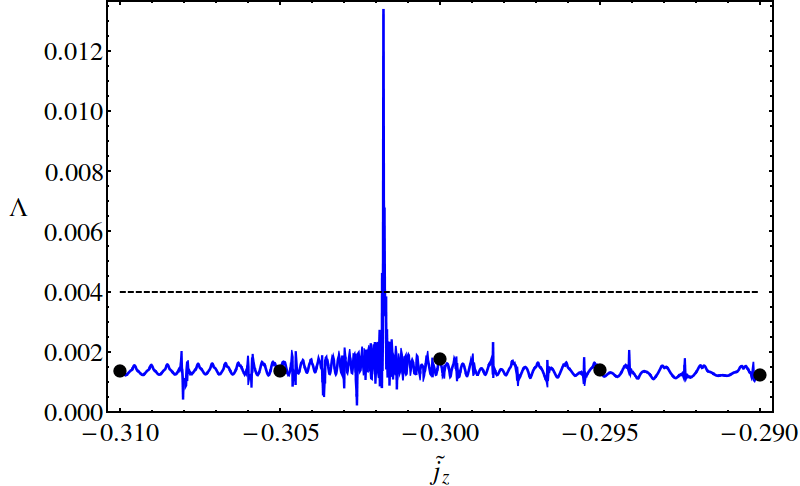}
\caption{Closer and detailed view of Fig.\ref{fig:6}(a), around the region $\tilde{j}_z=-0.003$. The zoom reveals the existence of chaotic trajectories in a very small interval around $\tilde{j}_z\approx -0.3018$, where the Lyapunov exponent takes values above the numerical tolerance $\Lambda_T=0.004$ (dashed line). The dots are  points considered  in  the Fig.\ref{fig:6}(a).}
\label{fig:7}
\end{figure}

For the set of parameters chosen, the ground-state energy is $\epsilon_{g.s.}=-2.125\, \omega_{0}$. Around the low energy value $\epsilon=-1.8 \, \omega_{0}$ we expect to have
a dominance of regular behavior. This is confirmed qualitatively by the regular orbits in the Poincar\'e sections as can be observed in the uppermost panels of figure \ref{fig:5}. The presence of low energy regular regions can be understood by the existence of approximate integrals of motion, associated with the adiabatic decoupling of the two collective degrees of freedom of the system \cite{BOA2016}. The existence of chaotic trajectories is not precluded, but they are  limited to the thin regions separating the sets of regular trajectories that can be observed in the Poincar\'e sections mentioned above. In Fig.\ref{fig:6} the Lyapunov exponent and the participation ratio (compared with the number of atoms) are pretty small, fully confirming that the region is predominantly regular. In the binary case, Fig. \ref{fig:6}(c,d), both quantities are null. The $P_{R}$ also  predicts the regularity of almost every point at the same extent as the Lyapunov exponent. However, a detailed and  closer view of  Fig.\ref{fig:6}(a)  around the region $\tilde{j}_z\approx-0.003$ (Fig.\ref{fig:7}) reveals the existence of chaotic trajectories in  a very small interval around  $\tilde{j}_z=-0.3018$,  where the Lyapunov exponent takes values well above the tolerance $\Lambda_T=0.004$, reaching $\Lambda\approx 0.013$. We will see in the next section that, even if this chaotic region is very small,  the $P_R$ is able to detect it when its  dependence on the number of atoms is considered.

\subsection{Mixed region $\epsilon=-1.5\,\omega_{0}$}

\begin{figure}
\centering
\begin{tabular}{cc}
(a)&(b)\\
\vspace{-1.4 cm}\\
\includegraphics[width=0.5\textwidth]{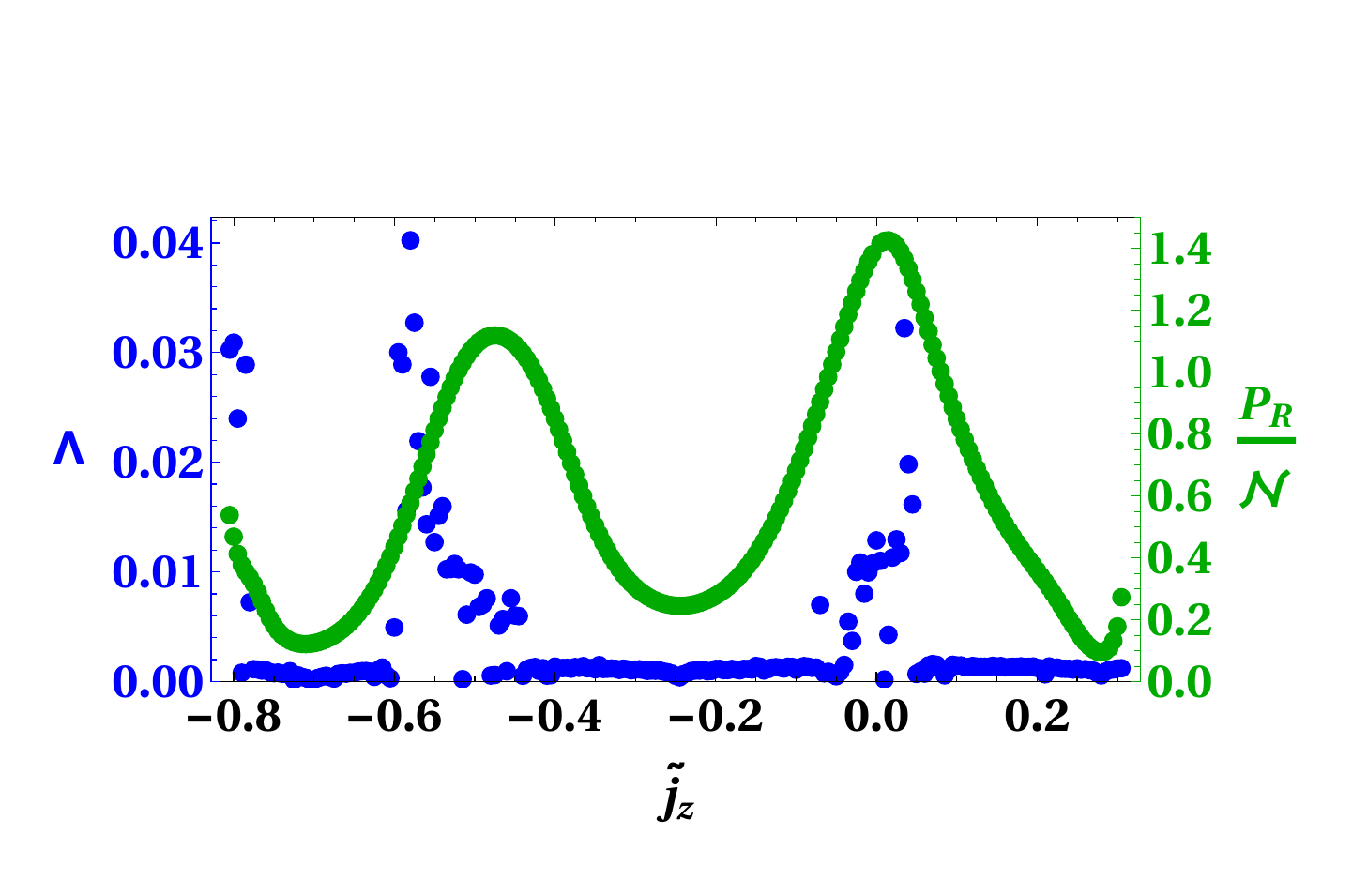} & \includegraphics[width=0.5\textwidth]{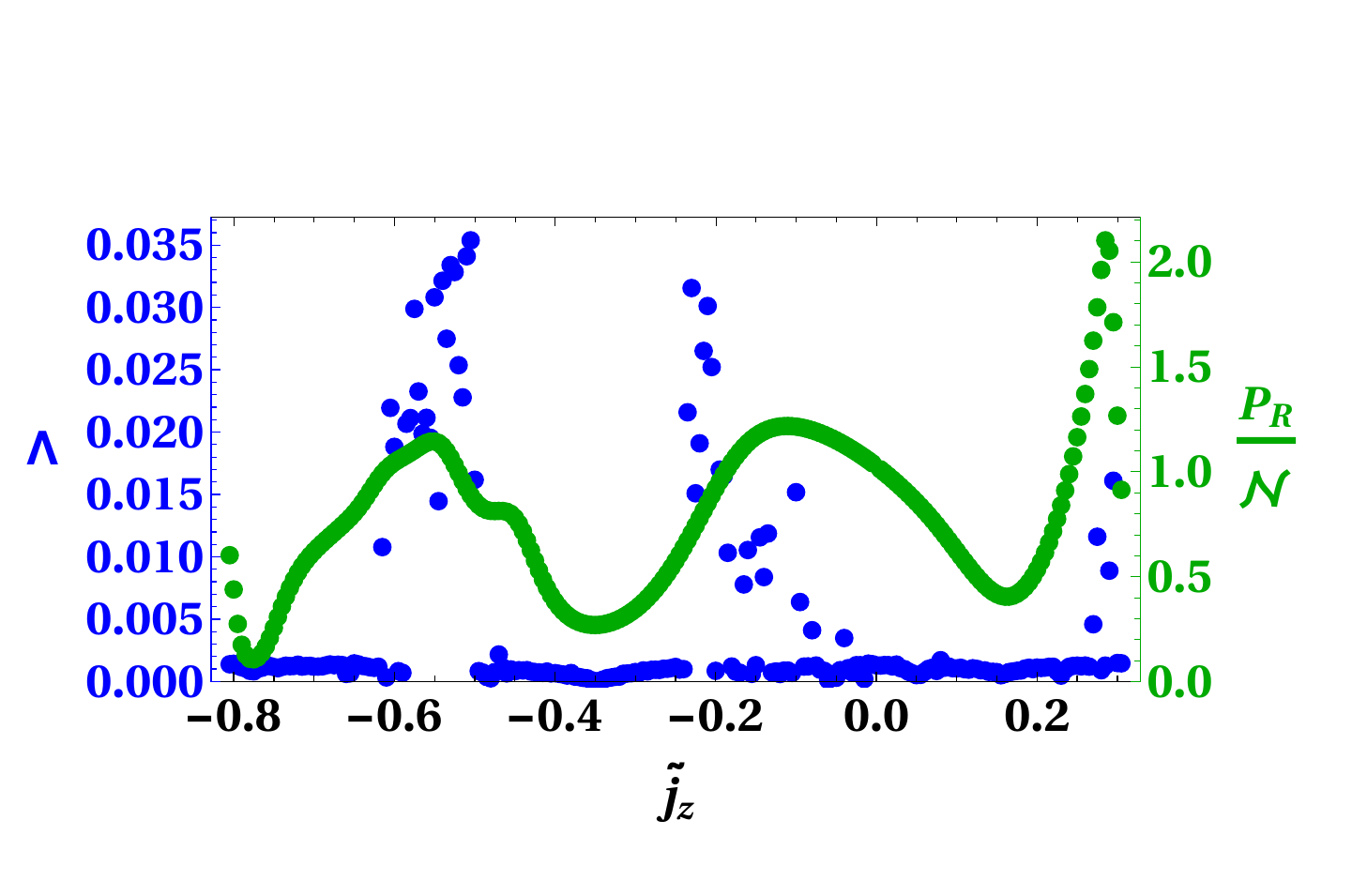} \\
(c)&(d)\\
\vspace{-1.4 cm}\\
\includegraphics[width=0.5\textwidth]{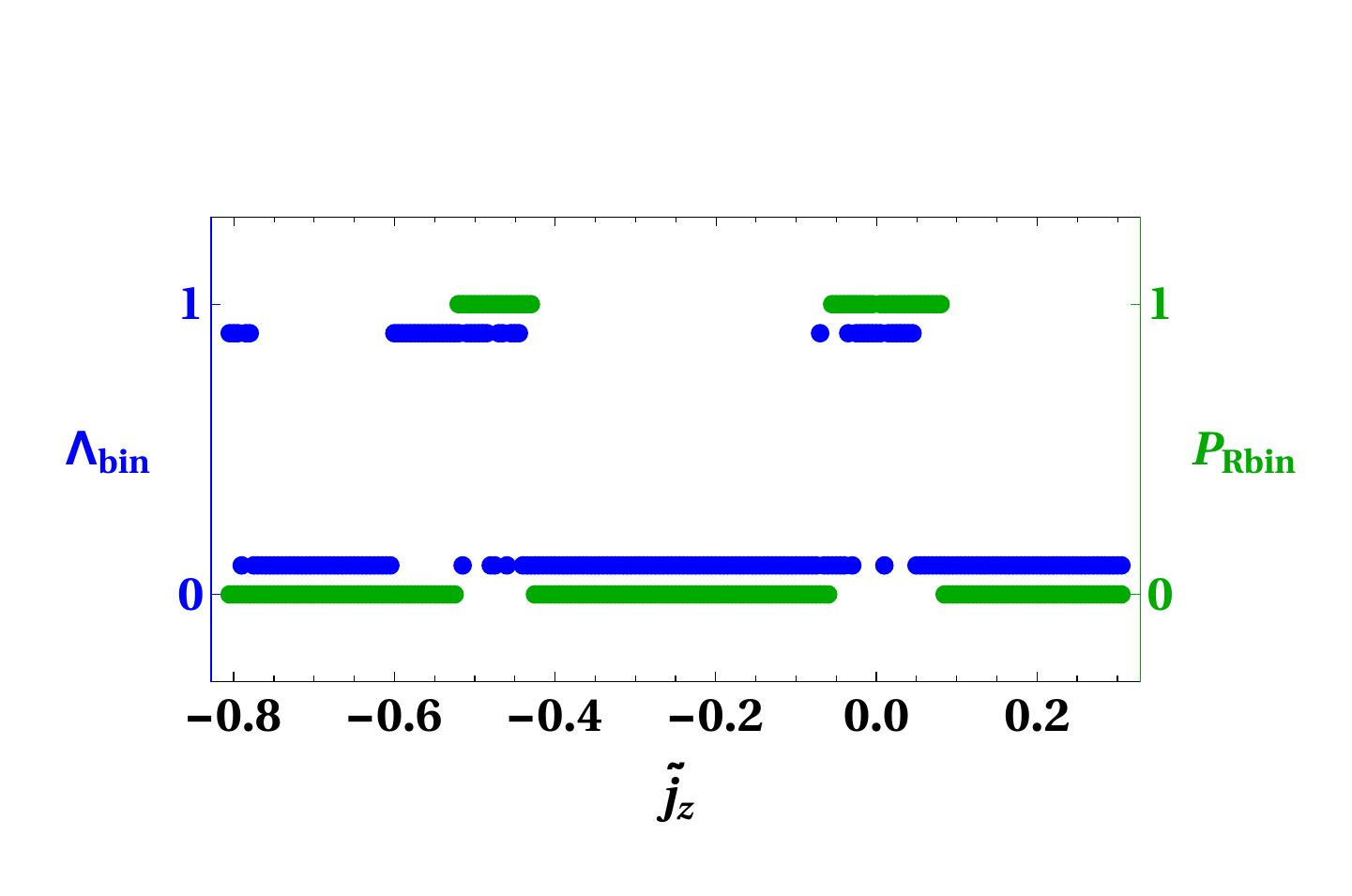} & \includegraphics[width=0.5\textwidth]{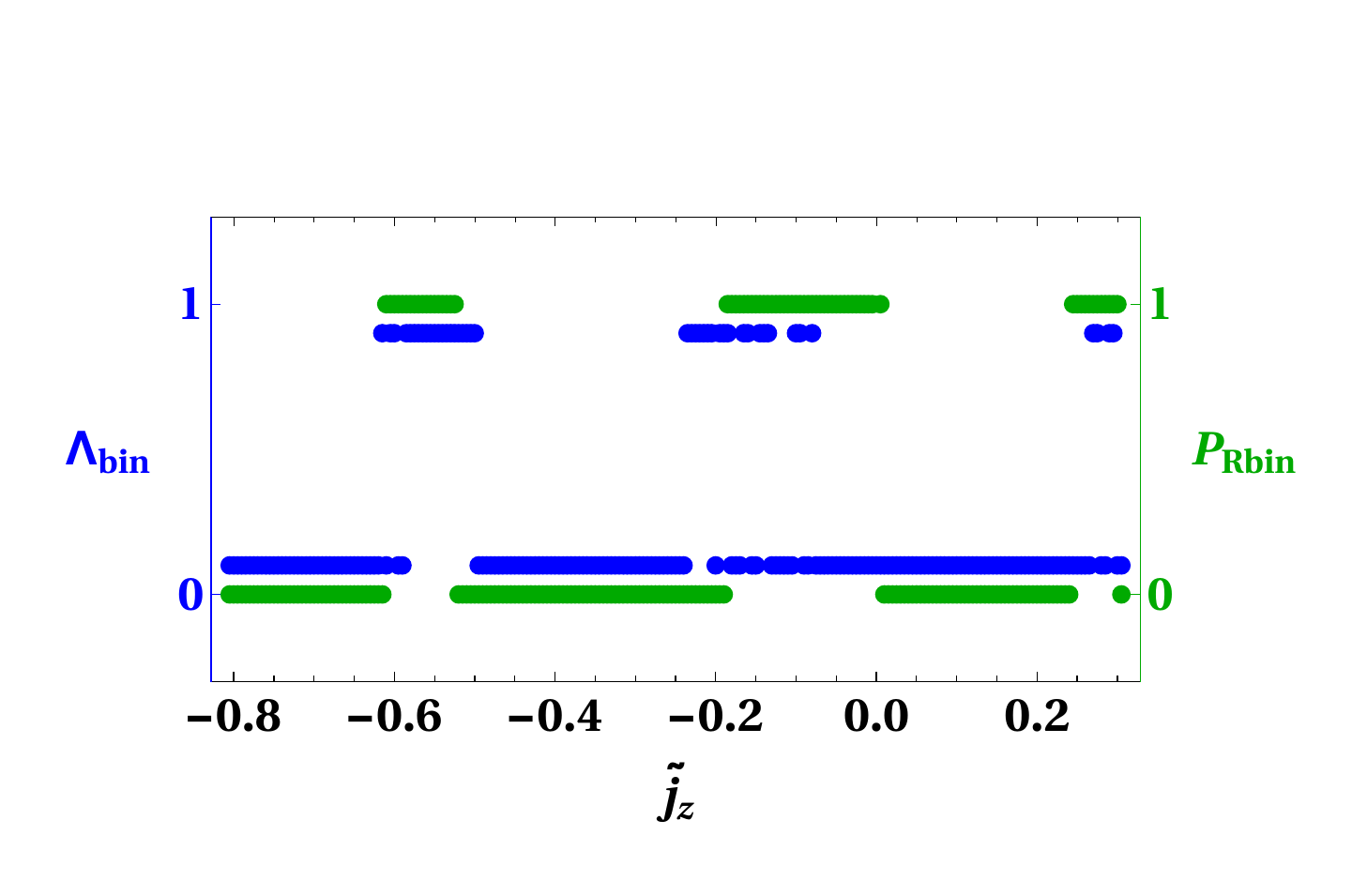} \\
\end{tabular} 
\caption{ Same as Fig.\ref{fig:6}, but for energy $\epsilon=-1.5\,\omega_{0}$, where  a mixed classical phase space is obtained, with regular and chaotic regions. This mixed phase space structure is reflected by the Lyapunov exponents, which are larger than $\Lambda_T=0.004$ in several intervals indicated in panel (c) and (d). These intervals are clearly correlated with those where the quantum $P_{R}/\mathcal{N}$ takes values larger than $1$.     
}
\label{fig:8}
\end{figure}

Moving up in energy to $\epsilon=-1.5\, \omega_{0}$, regularity and chaos coexist in the phase space as it can be seen from Fig.\ref{fig:8}, where there are three small but visible regions for each set of $q's$ where both the Lyapunov exponent and the participation ratio announce the presence of chaos. For $q_+$, Fig. \ref{fig:8} (a, c), they are around $\tilde{j_z}$ -0.8, -0.6 and 0.0. For $q_-$, Fig. \ref{fig:8} (b, d), they are around $\tilde{j_z}$ -0.6, -0.1 and 0.3. This is the most important point which we are communicating in this Comment:
{\em the quantum participation ratio ($P_R$) of coherent states on the eigenenergy basis plays a role equivalent to the Lyapunov exponent}. The support of this affirmation can be found in the figures of the present section. 

We are aware that the agreement between the regions in phase space described as chaotic employing the classical Lyapunov exponent  and the quantum participation ratio is not perfect. The differences are originated by two limitations. One is  the incertitude in the value of the tolerance cut which delimits regularity from chaos in the numerical studies, as mentioned in the previous section. The second one is related with the coarse grained nature of the quantum coherent states, which makes them unable to resolve structures smaller than the Planck cell, whose size, in the natural units used here,  is of the order  $2\pi/j$. Consequently, the participation ratio becomes a better identifier of the presence of chaos as the number of atoms ($\mathcal{N}=2j$) included in the calculation increases, but the numerical implementation can easily become prohibitive.   

\subsection{Just below the ESQPT region $\epsilon=-1.1\,\omega_{0}$}

\begin{figure}
\centering
\begin{tabular}{cc}
(a)&(b)\\
\vspace{-1.4 cm}\\
\includegraphics[width=0.5\textwidth]{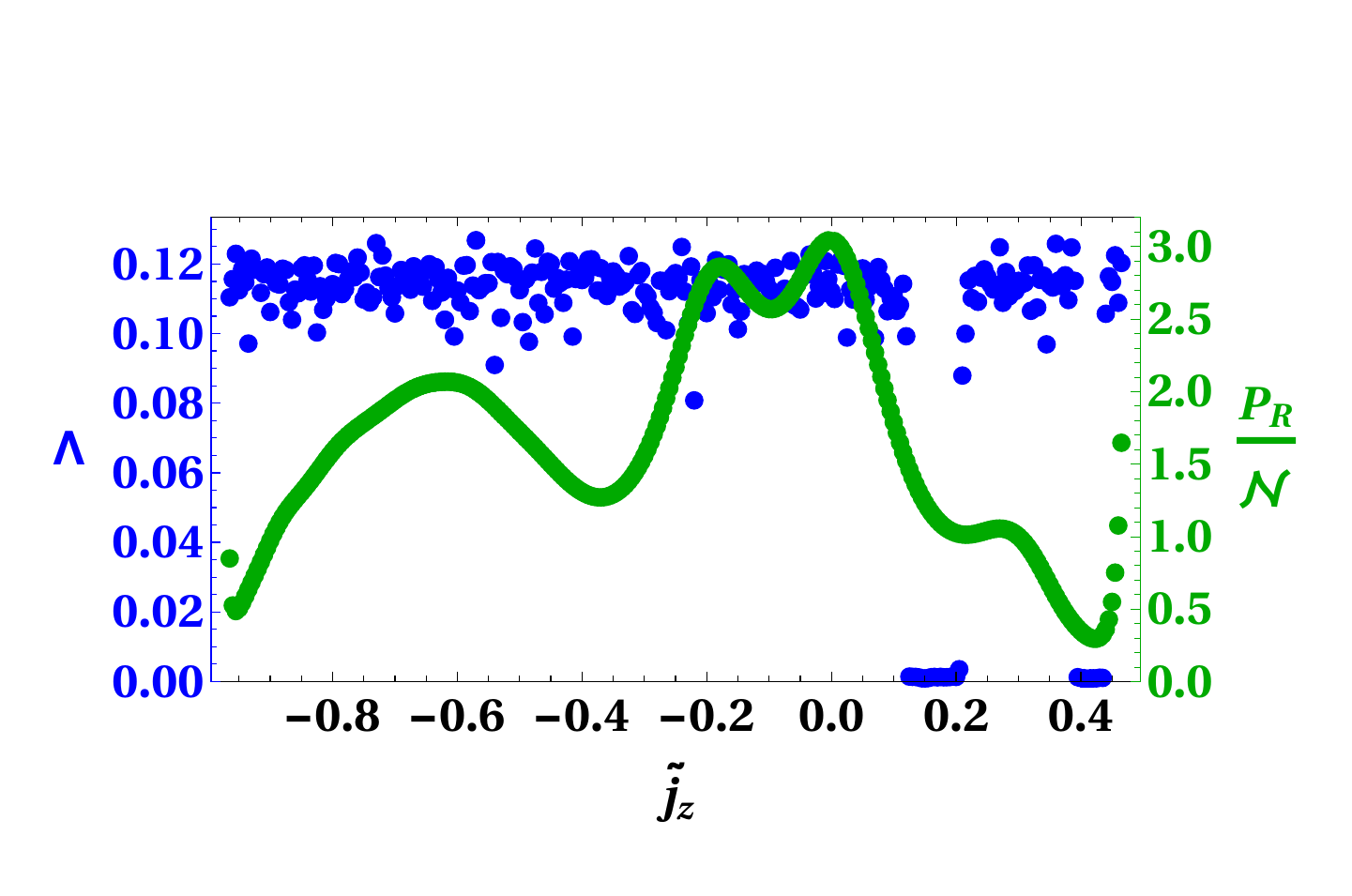} & \includegraphics[width=0.5\textwidth]{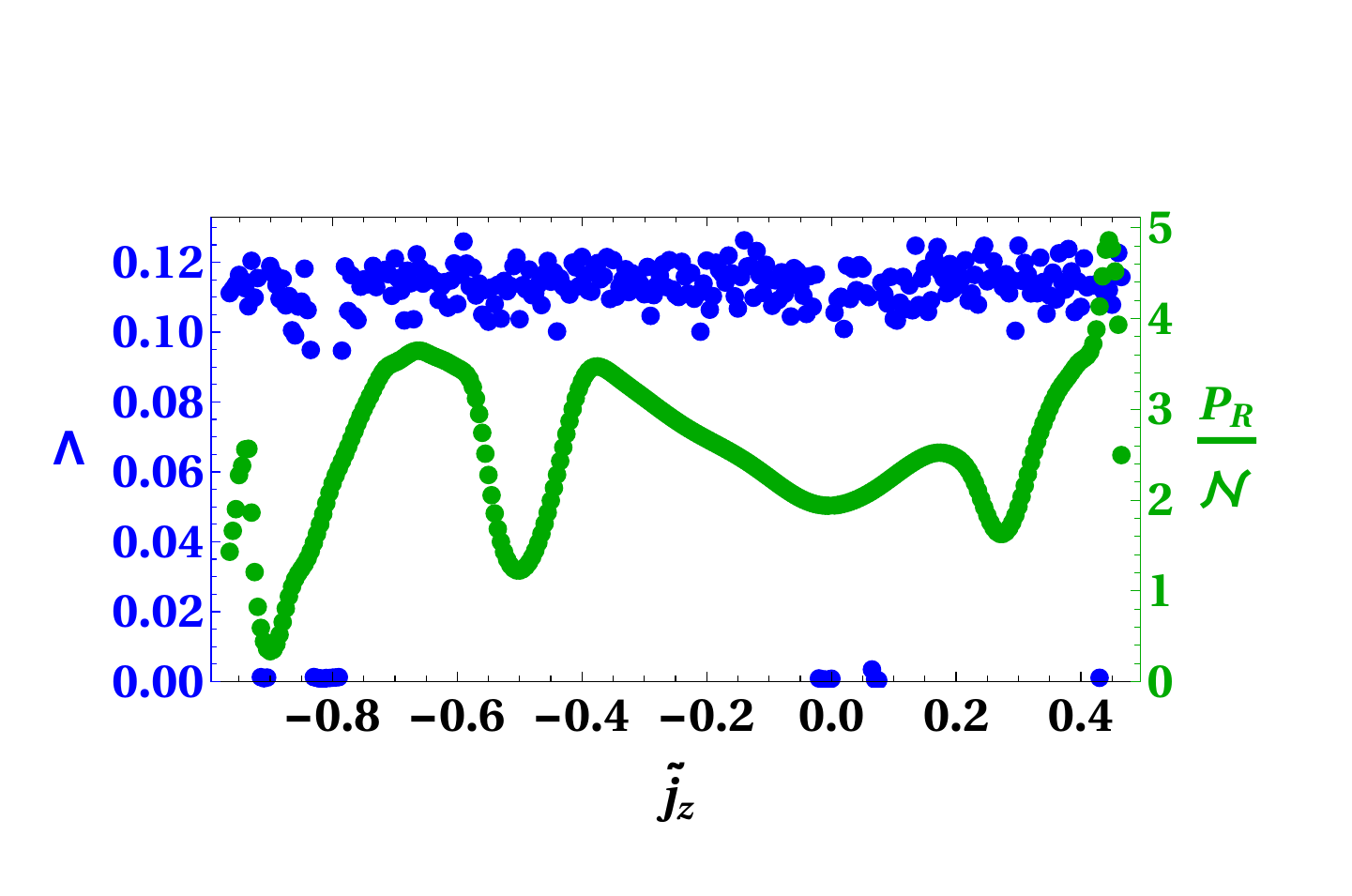} \\ 
(c)&(d)\\
\vspace{-1.4 cm}\\
\includegraphics[width=0.5\textwidth]{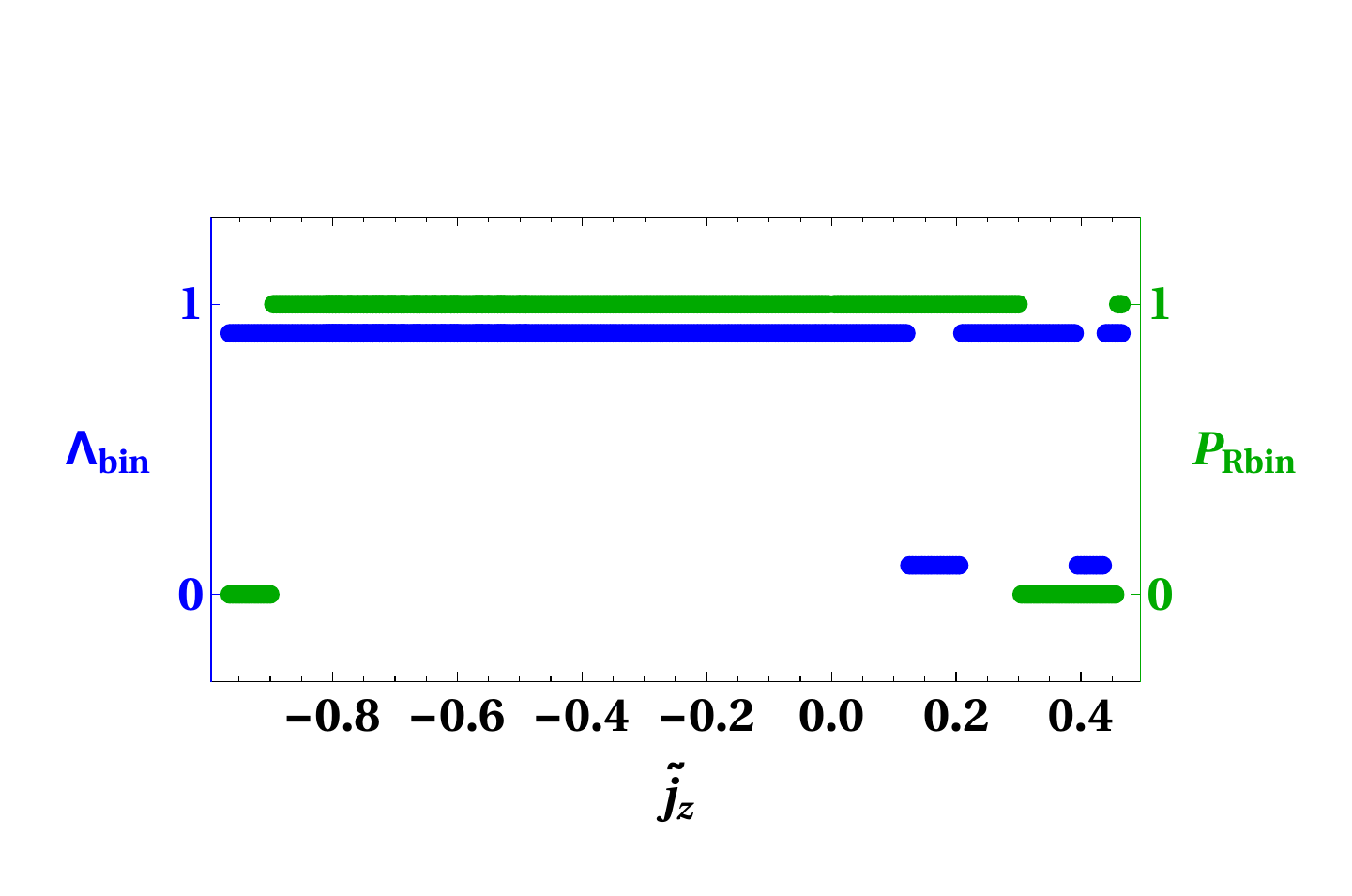} & \includegraphics[width=0.5\textwidth]{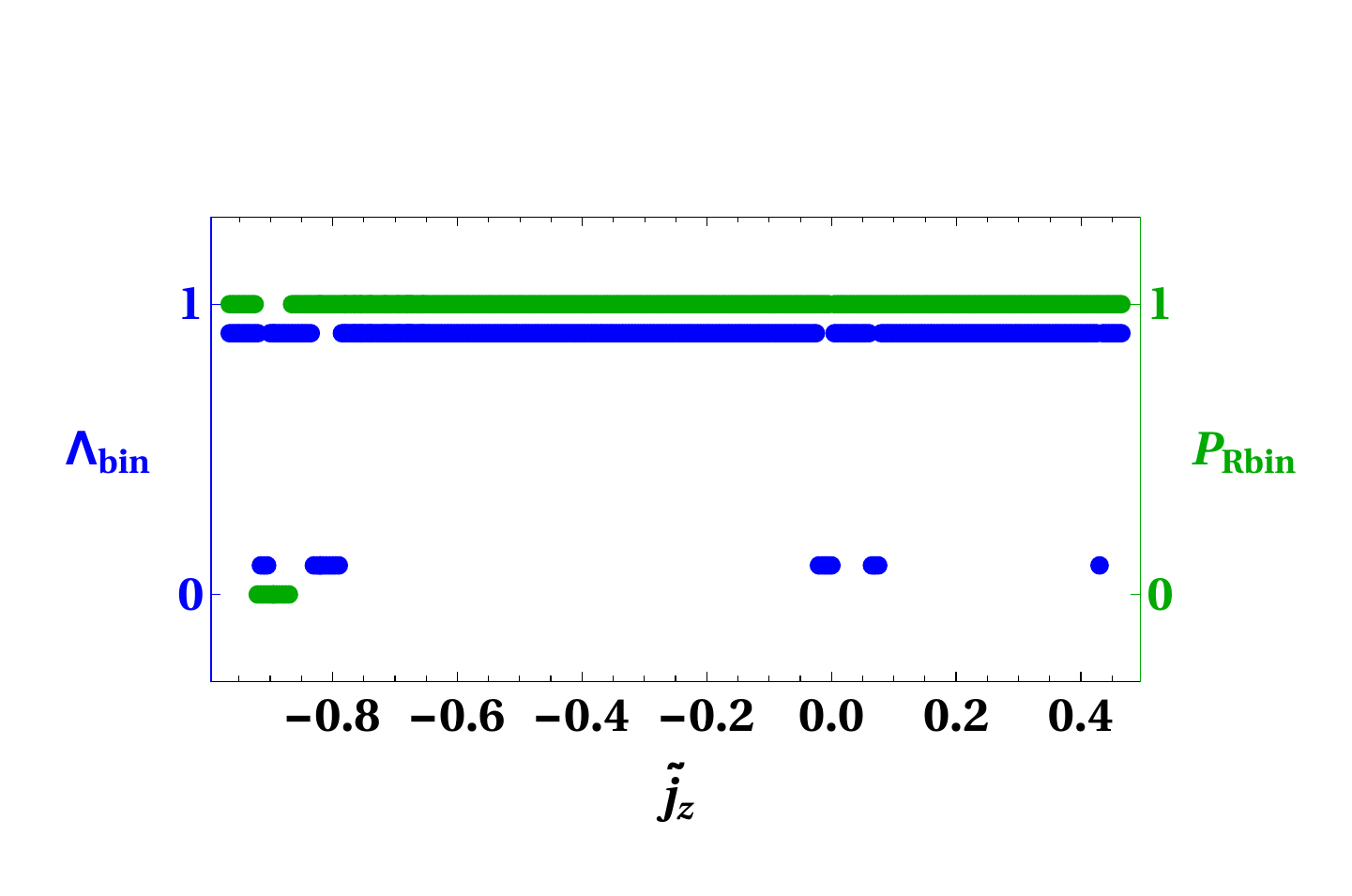} \\
\end{tabular} 
\caption{Same as Fig.\ref{fig:6}, but for energy $\epsilon=-1.1\,\omega_{0}$, just below the ESQPT critical energy ($\epsilon_c=-1$), where the classical phase space is almost entirely covered by chaotic trajectories. The presence of small island of stability is detected both by the Lyapunov and $P_{R}$.  
}
\label{fig:9}
\end{figure}

Moving upper in energy, at $\epsilon=-1.1\,\omega_{0}$ we approach the ESQPT from below. In this case  we observe in the phase space that the system is mostly chaotic, with some small islands of regularity. In Fig. \ref{fig:9} there are regions for each set of $q's$ where both the Lyapunov exponent and the participation ratio are still very small. For $q_+$, Fig. \ref{fig:9} (a,c) these regions  are around $\tilde{j_z}$  -0.9, 0.2 and 0.4. For $q_-$, Fig. \ref{fig:9} (b, d) they are around $\tilde{j_z}$ -0.9 and 0.0. Both  the Lyapunov exponent and the participation ratio detect these regions, with the limitations mentioned above.

\subsection{Just above the ESQPT region $\epsilon=-0.9\,\omega_{0}$}

\begin{figure}
\centering
\begin{tabular}{cc}
(a)&(b)\\
\vspace{-1.4 cm}\\
\includegraphics[width=0.5\textwidth]{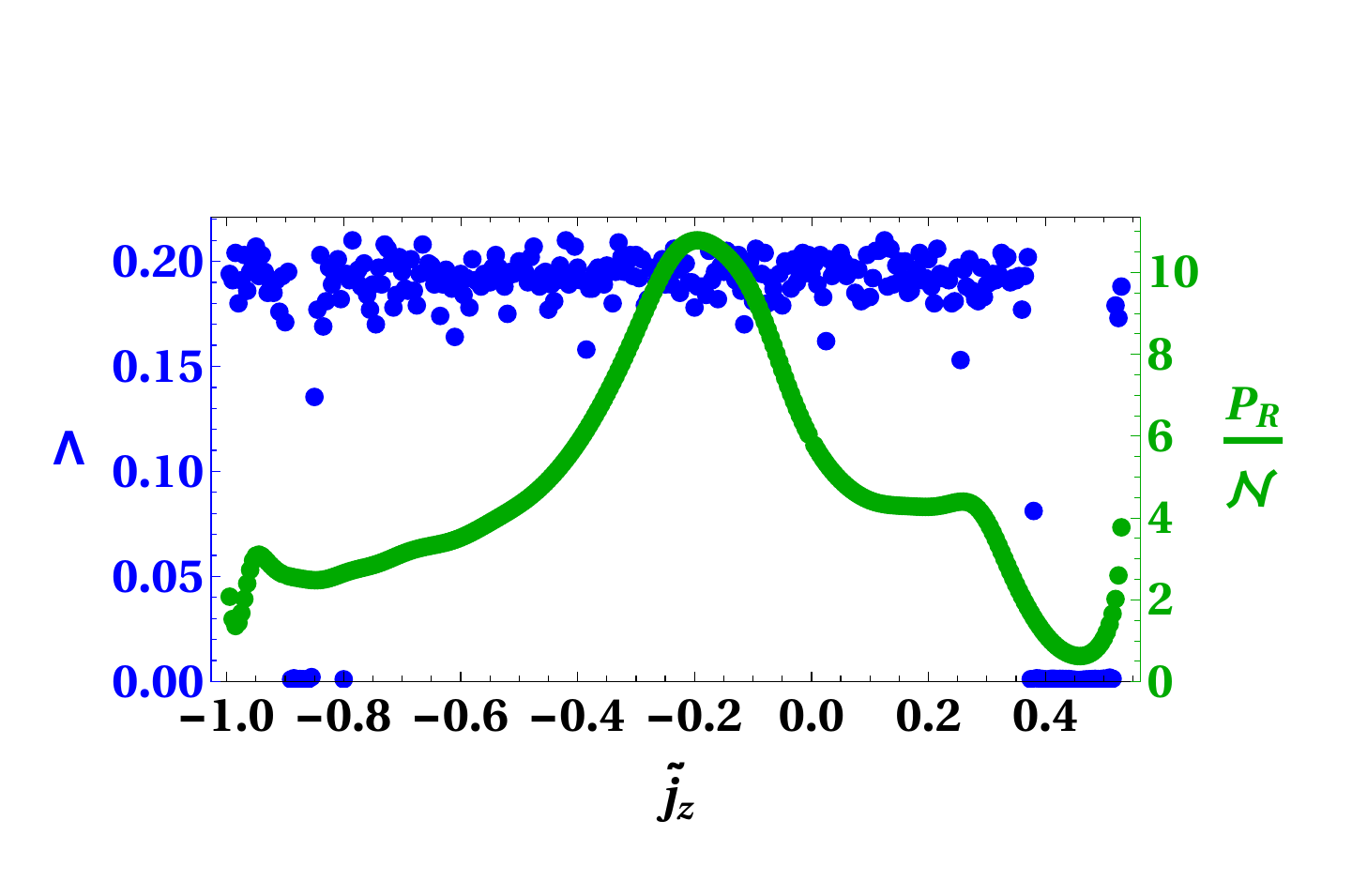} & \includegraphics[width=0.5\textwidth]{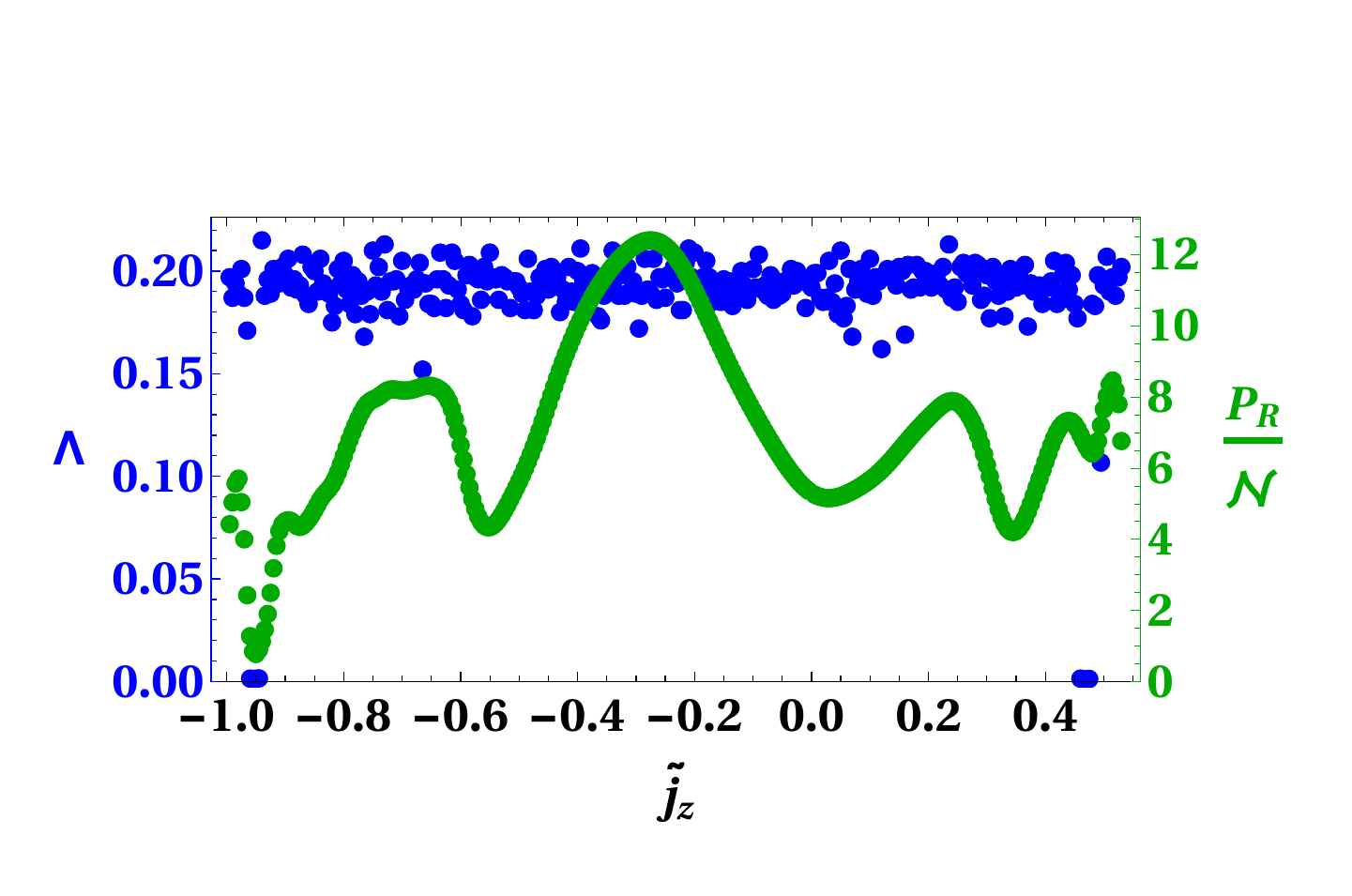}\\ 
(c)&(d)\\
\vspace{-1.4 cm}\\
\includegraphics[width=0.5\textwidth]{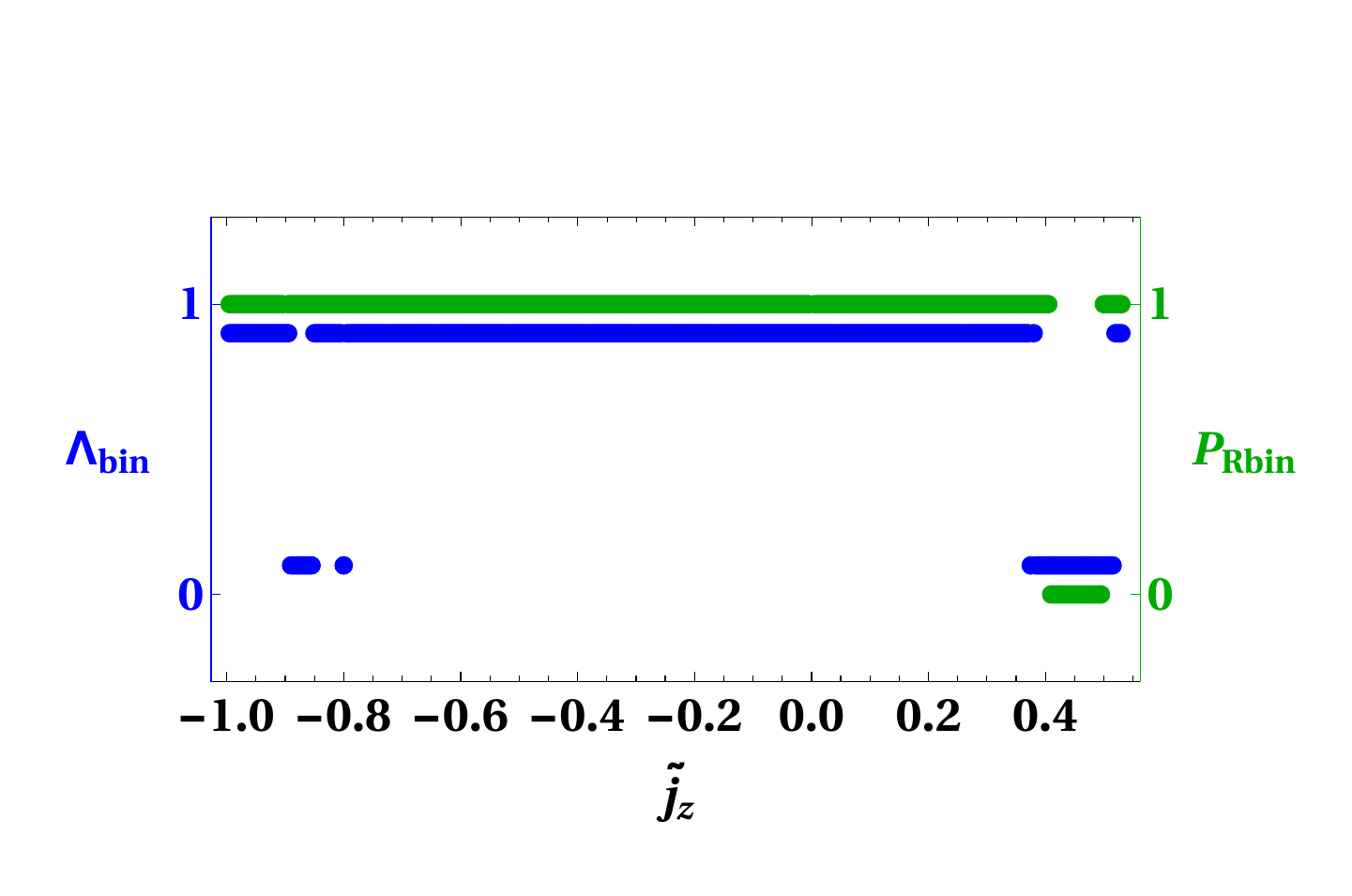} & \includegraphics[width=0.5\textwidth]{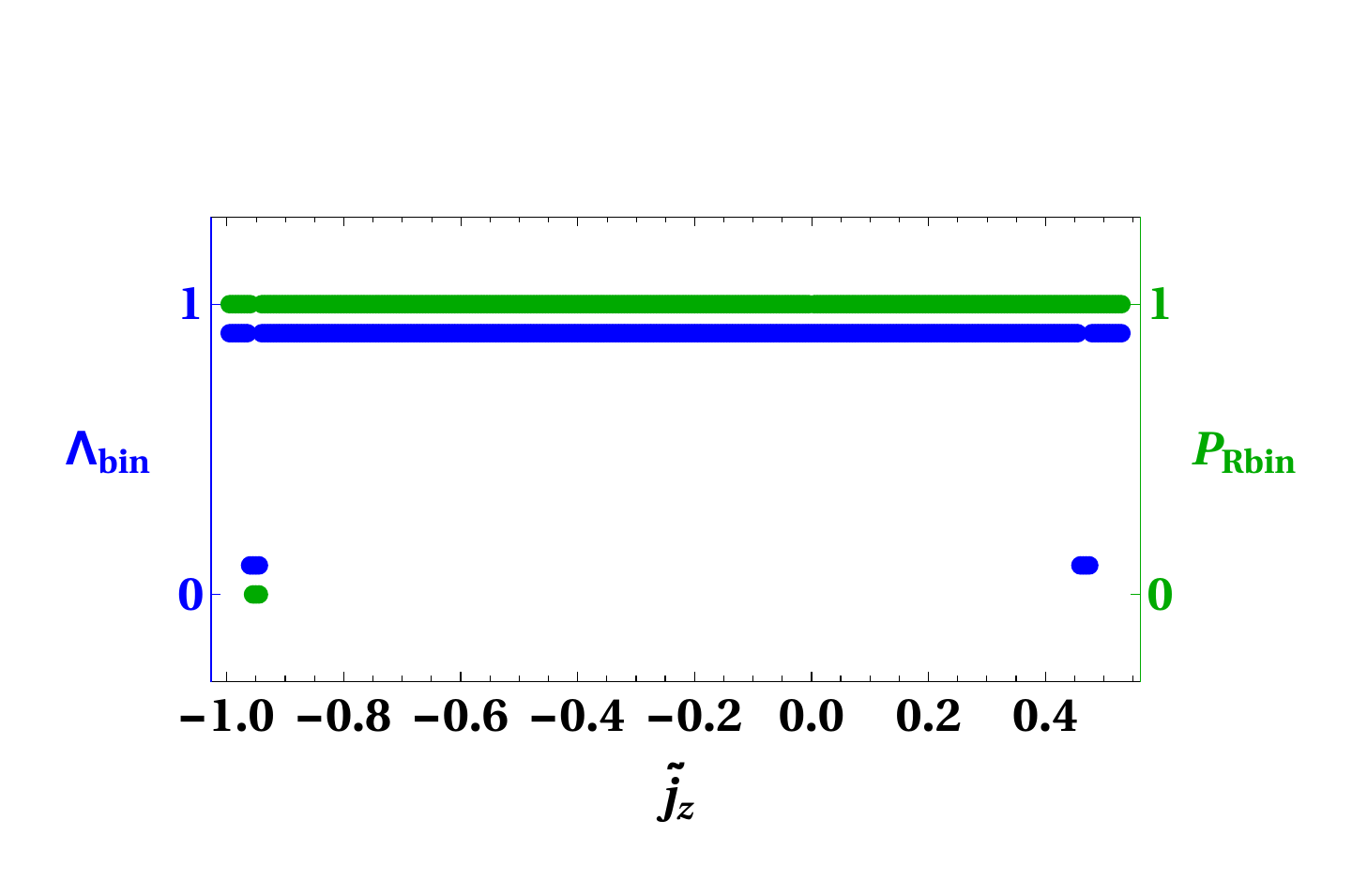} \\
\end{tabular} 
\caption{Same as Fig.\ref{fig:6}, but for energy $\epsilon=-0.9\,\omega_{0}$, just above the ESQPT critical energy,  where the classical phase space is, similar to the previous figure, almost entirely covered by chaotic trajectories. As in previous figure, both the Luapunov and $P_{R}/\mathcal{N}$ are able to detect the small islands of stability   
}
\label{fig:10}
\end{figure}

For an energy $\epsilon=-0.9\,\omega_{0}$ we are located just above  the ESQPT. As in the previous case,  the system is almost fully chaotic, with some small islands of regularity. In Fig.\ref{fig:10} there are just a couple of regions for each set of $q's$ where both the Lyapunov exponent and the participation ratio are close to zero. For $q_+$, Fig.\ref{fig:10} (a, c), that couple is around $\tilde{j_z}$ -0.9 and 0.4. For $q_-$, Fig.\ref{fig:10} (b, d) that is around $\tilde{j_z}$  -1.0 and 0.5. Again, both the Lyapunov exponent and the participation ratio detect these regions.

\subsection{Fully chaotic region $\epsilon=-0.5\,\omega_{0}$}

\begin{figure}
\centering
\begin{tabular}{cc}
(a)&(b)\\
\vspace{-1.4 cm}\\
\includegraphics[width=0.5\textwidth]{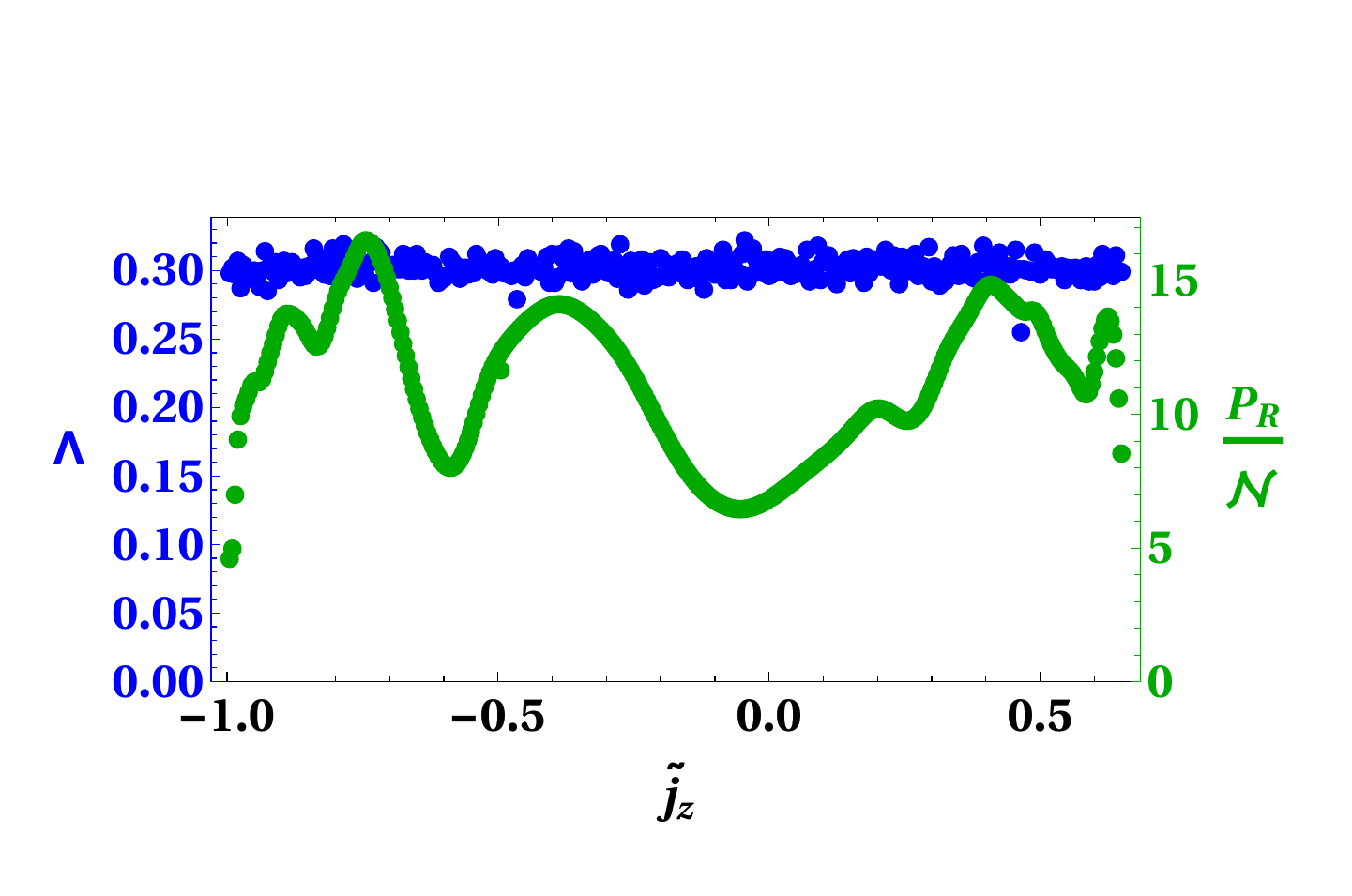} & \includegraphics[width=0.5\textwidth]{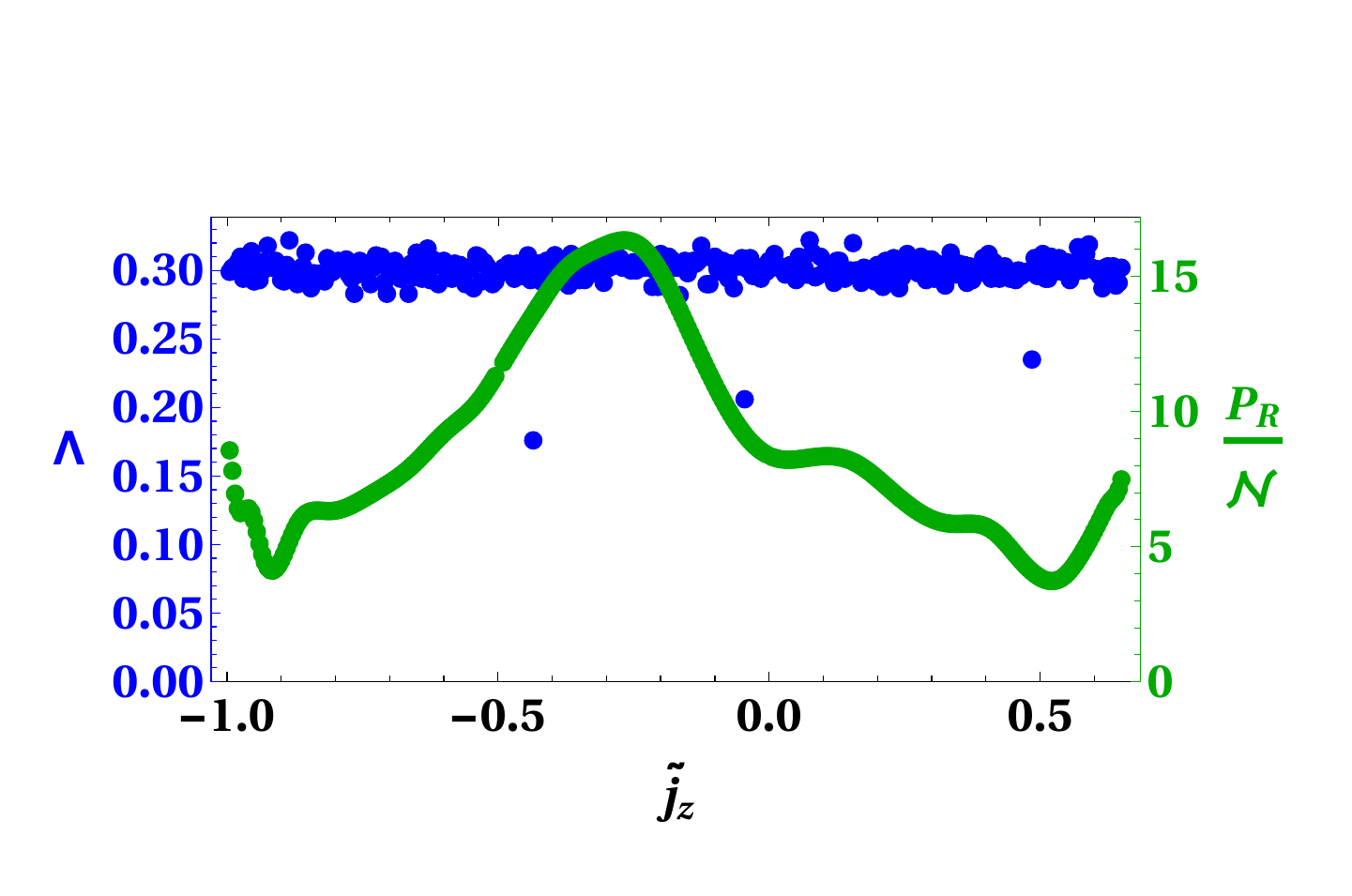} \\ 
(c)&(d)\\
\vspace{-1.4 cm}\\
\includegraphics[width=0.5\textwidth]{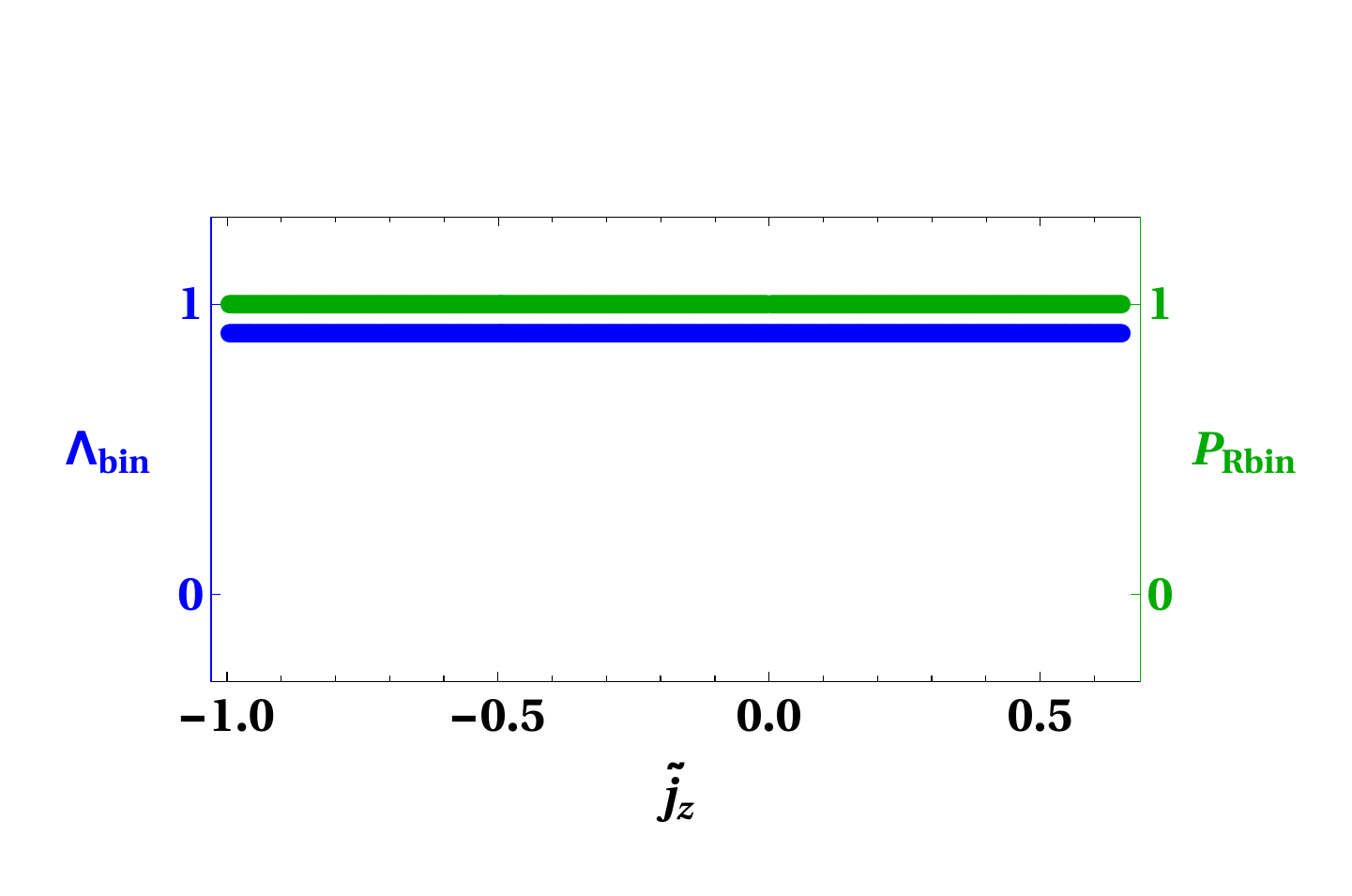} & \includegraphics[width=0.5\textwidth]{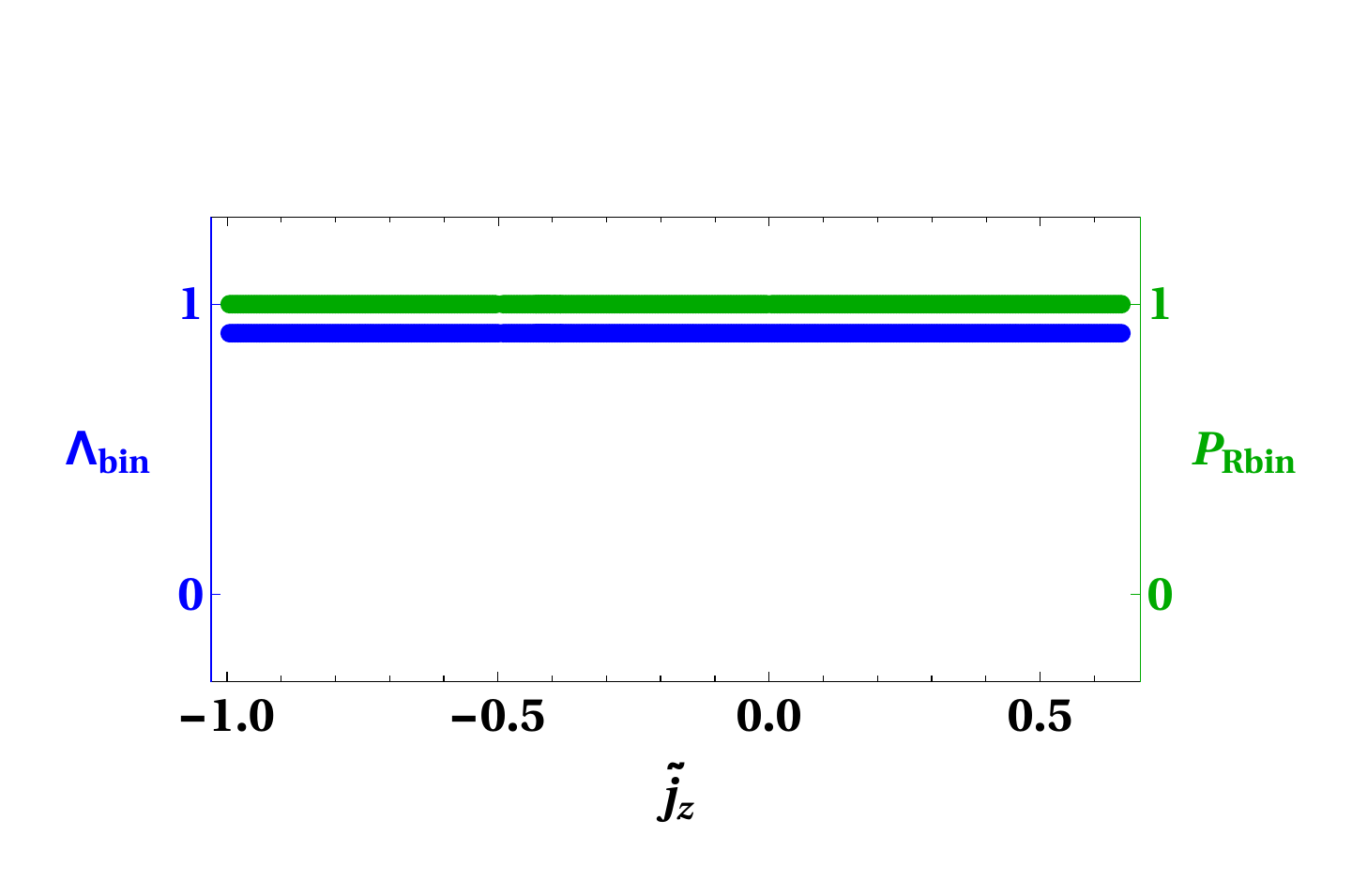} \\
\end{tabular} 
\caption{Same as Fig.\ref{fig:6}, but for  energy $\epsilon=-0.5\,\omega_{0}$, well above the ESQPT critical energy, where the ergodic regime has been reached.  This ergodicity is  reflected by the Lyapunov and $P_{R}$, which only  take  values well above the tolerances employed to determine chaotic dynamics ($\Lambda>\Lambda_T=0.004$ and $P_{R}/\mathcal{N}>1$ respectively).    
}
\label{fig:11}
\end{figure}

At  energy $\epsilon=-0.5\,\omega_{0}$ the phase space is  fully chaotic. In Fig.\ref{fig:11} both the Lyapunov exponent and the participation ratio have values far larger than their numerical tolerance cuts. The comparison between them is straightforward, as it was for the regular  cases  at low energies. Both criteria indicates the phase space is completely  chaotic. 

As mentioned above, in all cases we observe a very good global agreement between the $P_{R}$ and the Lyapunov exponent. The global agreement is better seen in the case of the binary criterion. The differences that remain can be attributed to the finite value of $j$. In the thermodynamical limit we expect the Lyapunov and the $P_{R}$  binary quantities  will perfectly match.

\section{Scaling of the $P_{R}$}

We have already exhibited that the behavior of the classical Lyapunov exponent and the quantum participation ratio follow each other, they are both large in chaotic regions and small in the regular ones. We have also corroborated that the  limits for the binary quantities, $\Lambda_{bin}=0$ if $\Lambda\leq 0.004$ and $P_{Rbin}=0$ if $P_{R} \leq \mathcal{N}$, define a numerical simple criteria to distinguish regularity or chaos. However,  as we advanced in Ref. \cite{Basta16PRE},  a more precise quantitative criteria to determine if a particular point in phase space is associated with a regular or chaotic dynamics, involves the numerically more demanding task of evaluating the $P_{R}$ as a function of $\mathcal{N}=2j$, and analyzing its functional dependence. If $P_R\propto \mathcal{N}^{\alpha}$ with $\alpha<1$ then $\lim_{j\rightarrow\infty}{P_{R}}/{\mathcal{N}}$ goes to zero, and  we can confidently conclude that this coherent state is localized in the Hamiltonian eigenstate basis, and that the associated classical trajectory is regular. If this condition  is not fulfilled ($P_R\propto \mathcal{N}^{\alpha}$ with $\alpha\geq 1$), this point in phase space has a chaotic dynamics.   

\begin{figure}
\centering
\begin{tabular}{ccc}
(a)&(b)&(c)\\
\vspace{-0.5cm}\\
\includegraphics[width=0.33\textwidth]{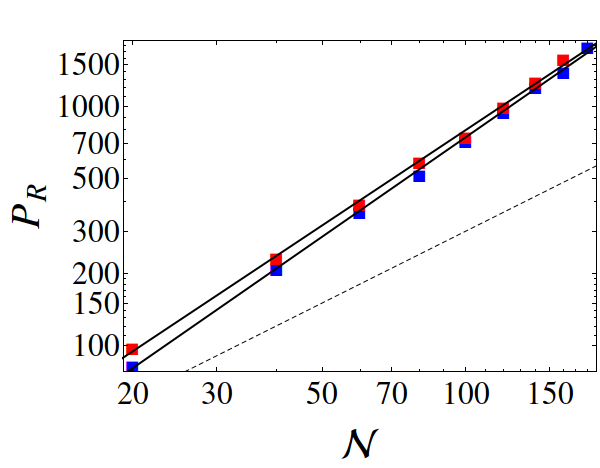}&\includegraphics[width=0.33\textwidth]{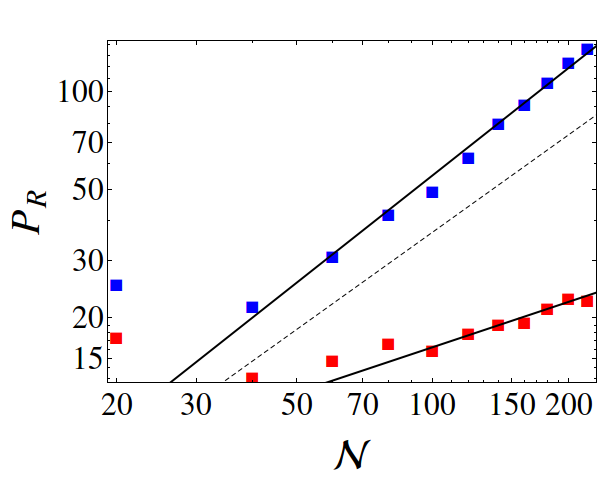}&\includegraphics[width=0.33\textwidth]{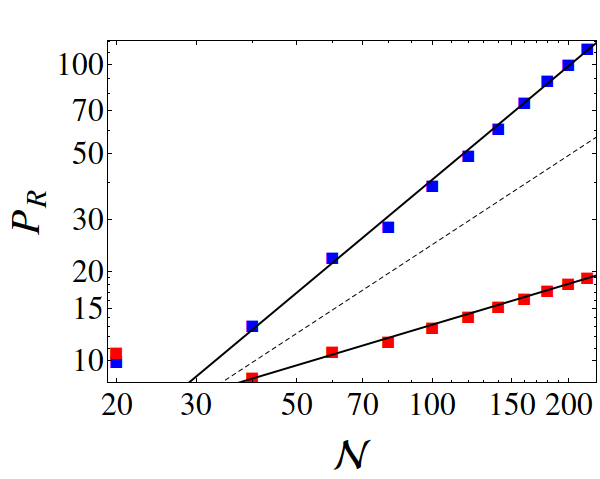}
\end{tabular} 
\caption{The participation ratio $P_R$ as a function of the number of atoms $\mathcal{N}$ in log-log scale, for three different energies $\epsilon=-0.5\,\omega_{0}$ (a), $\epsilon=-1.5\,\omega_{0}$ (b), and $\epsilon=-1.8\,\omega_{0}$ (c). The blue (red) dots represent the values for the points ($\phi=0$,  $p=0$, $\tilde{j}_z$ and   $q_+$) with the largest (smallest) Lyapunov exponent in  panels (a) of Figs.\ref{fig:6}, \ref{fig:8} and \ref{fig:11}. The continuous lines correspond to the fits listed in Table \ref{table1}. The dashed lines show the slope of a linear dependence $\propto N$, giving a guide to distinguish between  scalings faster or slower than the number of atoms.}
\label{fig:12}
\end{figure}

In Fig \ref{fig:12} we show the scaling of some specific points which belong to the energy surfaces explored above. We have selected two points in phase space for  three energies, $\epsilon=-0.5 \, \omega_{0}$ (a), $\epsilon=-1.5\, \omega_{0}$ (b), and $\epsilon=-1.8 \, \omega_{0}$ (c).  The selected points in each case correspond to those with the largest  (blue) and smallest (red) Lyapunov exponent in the $q_+$ branch of the energy surface. They are listed in Table \ref{table1}, where the values of the energy $\epsilon$, the pseudospin coordinate $\tilde{j}_{z}$, the Lyapunov exponent  $\Lambda$ and the function providing a fit of the data are given. The fitting functions are representative of the asymptotic behaviour of the participation ratios for large number of atoms. They scale as $P_R\propto N^\alpha$ with $\alpha\approx 1/2$ for regular points and as $P_R\propto N^\alpha$ with $\alpha\gtrsim 1$ for chaotic ones.

\begin{table}
\centering
\begin{tabular}{|c|c|c|c|} \hline
 & & & \\
$\epsilon/\omega_{0}$ & $\tilde{j}_{z}$ & $\Lambda$ & Fit \\ \hline
 & & & \\
-0.5 & -0.435 & 0.176 & $P_{R}=1.76\mathcal{N}^{1.328}$ \\ \hline
 & & & \\
-0.5 & 0.075 & 0.322 & $P_{R}=1.26 \mathcal{N}^{1.386}$\\ \hline
 & & & \\
-1.5 & -0.300 &0.00014 & $P_{R}=0.64\mathcal{N}^{0.466}$ \\ \hline
 & & & \\
-1.5 & -0.580 & 0.040 & $P_{R}=0.34\mathcal{N}^{1.103}$ \\ \hline
 & & & \\
-1.8 & -0.495 & $9.18 \times 10^{-6}$& $P_{R}=1.61 \mathcal{N}^{0.458}$ \\ \hline
 & & & \\
-1.8 & -0.3018 &  0.013   & $P_{R}=0.12\mathcal{N}^{1.274}$  \\ \hline
\end{tabular}
\caption{The six points in phase phase for which the scaling of the participation ratio with the number of atoms is plotted in Fig. \ref{fig:12}. A Lyapunov expnent $\Lambda>\Lambda_T=0.004$ indicates a chaotic classical trajectory. }
\label{table1}
\end{table}

At the highest energy $\epsilon=-0.5$, Fig. \ref{fig:12} (a), it can be seen how the participation ratio grows faster than the number of atoms, revealing both points as  chaotic, in correspondence with the Lyapunov exponent which is much larger than $0.004$ for both points. At the intermediate energy $\epsilon=-1.5$, Fig. \ref{fig:12} (b), the participation ratio  associated with the largest Lyapunov exponent (much greater than $\Lambda_T=0.004$) shown as blue points, grows faster than the number of atoms and is identified also as chaotic. On the other hand, the red dots, representing the participation ratio of the point with the smallest Lyapunov exponent (much smaller than $\Lambda_T=0.004$), scales much slower than $\mathcal{N}$, allowing its identification as a regular point in phase space. 
At the low energy $\epsilon=-1.8$, Fig. \ref{fig:12} (c), a pattern similar to the one described for the intermediate energy  occurs. For the point with the smallest Lyapunov exponent (much smaller than the tolerance $\Lambda_T=0.004$), we find, as expected,  a scaling $P_R\propto N^{1/2}$.
It is remarkable that in the case with the largest Lyapunov exponent, even if the chaotic classical region is very small (see Fig.\ref{fig:7}), the scaling of the participation ratio is faster than  $\mathcal{N}$, which is consistent with a chaotic dynamics. 
 This last case  exhibits the dependence of $P_R$ on  $\mathcal{N}$ as a very sensitive test, which is able to identify chaotic points in a phase space  which is almost 100\% regular (see uppermost panels of Fig.\ref{fig:5}). 

\section{Conclusions}

In this contribution we have exhibited the participation ratio $P_R$ of coherent states on the eigenenergy basis as a purely quantum tool to identify the regular and chaotic regions in phase space for systems which have an algebraic Hamiltonian, in the particular case of the Dicke Hamiltonian. It requires the introduction of  coherent states, which provides the bridge connecting the quantum and classical realms.  This work goes far beyond the qualitative analysis based in Poincar\'e sections and Peres lattices presented in \cite{Basta15PS}, explotes quantitatively the semiclassical study on chaos in the Dicke model discussed in  \cite{Cha16}, and extends the results discussed in \cite{Basta16PRE}.    

As the non-integrable Dicke model has regions associated with regular and chaotic motion, we have quantitatively evaluated the presence of classical chaos employing the largest Lyapunov exponent in the whole available phase space for a given energy.  In the quantum regime, the use of very efficient diagonalization techniques played a key role allowing the numerical evaluation of the quantum participation ratio $P_{R}$ of coherent states on the eigenenergy basis. We have shown that it plays a role equivalent to the Lyapunov exponent. We have also shown that, in the thermodynamic limit, the quotient of the participation ratio and the number of atoms  goes to zero in the regular regions and tends to a positive value in the chaotic ones, thus allowing us to define a criterion to identify chaos employing only quantum concepts. The $P_{R}$ is sensitive enough to detect regions with chaos on the phase space of almost regular regimes, restricted to the size of the Planck cell. 

The present proposal can be applied to any quantum mechanical system when it is possible to build a classical phase space through a semiclassical treatment, and  it  can be useful for other algebraic non-integrable models. 

\section*{Acknowledgments} We acknowledge financial support from Mexican CONACyT projects CB166302, CB2015-01/255702 and DGAPA-UNAM project IN109417. S.L-H. acknowledges financial support from the CONACyT fellowship program for sabbatical leaves.


\end{document}